\documentclass[graybox]{svmult}

\usepackage{type1cm}        
\usepackage{makeidx}         
\usepackage{graphicx}        
\usepackage{multicol}        
\usepackage[bottom]{footmisc}

\usepackage{newtxtext}       
\usepackage[varvw]{newtxmath}       

\usepackage{comment}

\makeindex             

\begin{document}

\title*{Impact of primordial black holes on the formation of the first stars and galaxies}
\author{Boyuan Liu and Volker Bromm}
\institute{Boyuan Liu \at Institute of Astronomy, University of Cambridge, Madingley Road, Cambridge, CB3 0HA, UK \email{bl527@cam.ac.uk}
\and
Volker Bromm \at Department of Astronomy, University of Texas, Austin, TX 78712, USA, \email{vbromm@astro.as.utexas.edu}}
%
%
\maketitle

\abstract{Recent GW observations of binary BH mergers and the SGWB have triggered renewed interest in PBHs in the stellar-mass ($\sim 10 - 100\ \rm M_\odot$) and supermassive regimes ($\sim 10^7 - 10^{11}\ \rm M_\odot$). Although only a small fraction ($\lesssim 1\%$) of DM in the form of PBHs is required to explain such observations, these PBHs may play important roles in early structure/star/galaxy formation. In this chapter, we combine semi-analytical analysis and cosmological simulations to explore the possible impact of PBHs on the formation of the first stars and galaxies, taking into account two (competing) effects of PBHs: acceleration of structure formation and gas heating by BH accretion feedback. We find that the impact of stellar-mass PBHs (allowed by existing observational constraints) on primordial star formation is likely minor, although they do alter the properties of the first star-forming halos/clouds and can potentially trigger the formation of massive BHs, while supermassive PBHs serve as seeds of massive structures that can explain the apparent overabundance of massive galaxies in recent observations. Our tentative models and results call for future studies with improved modeling of the interactions between PBHs, particle DM, and baryons to better understand the impact of PBHs on early star/galaxy/structure formation and their imprints in high-redshift observations.
}

\section{Introduction}
\label{sec:intro}

Recent GW observations provide novel hints on the presence of PBHs \cite{Hawking:1974rv,Carr:1974nx} as an important component of DM. In particular, mergers of stellar-mass ($\sim$10$ - 100\,\rm M_\odot$) PBHs can potentially explain the entire population of the binary black hole (BBH) mergers detected by the LIGO-Virgo-KAGRA Scientific Collaboration \cite{LIGOScientific:2020kqk}, if such PBHs make up a small mass fraction of DM $f_{\rm PBH}\sim 0.1-1\%$ \cite{Ali-Haimoud:2017rtz,Raidal:2017mfl,Hutsi:2020sol,Wong:2020yig,Andres-Carcasona:2024wqk}. Although it is unlikely that all BBH mergers originate from PBHs \cite{Hall:2020daa}, a non-negligible contribution from PBH mergers (with {$f_{\rm PBH}\sim 0.01-0.1\%$}) is favored by the observed rate of mass-gap events\footnote{Here, `mass gap' refers to the mass range ($\sim50-130\ \rm M_\odot$) unachievable for a remnant black hole within standard stellar evolution theory \cite{Mapelli:2021taw}.} like GW190521 {\cite{DeLuca:2021wjr,Franciolini:2021tla} (see also Part IV and Chapter 25)}. On the other hand, the SGWB recently detected by pulsar timing arrays \cite{NANOGrav:2023hfp} can also be explained by mergers of supermassive ($\sim 10^{7}-10^{11}\,\rm M_{\odot}$) PBHs given a similar PBH mass fraction in DM of {$f_{\rm PBH}\sim 0.1-1\%$ \cite{Gouttenoire:2023nzr,Huang2023}, although such a high abundance of supermassive PBHs (SMPBHs) is (marginally) in tension with local observational constraints that require $f_{\rm PBH}\lesssim 10^{-3}$ (independent of PBH formation mechanisms) \cite{Carr:2020gox}, even if we ignore the high-$z$ CMB $\mu$-distortion constraint that rules out PBHs with $m_{\rm PBH}\gtrsim 10^{4}\ \rm M_\odot$ formed by the spherical collapse of overdensities in a Gaussian density field} (see Sec.~\ref{sec:smpbh} and Part V of this book). Moreover, even if PBHs in these two mass regimes only account for a small fraction of DM, their existence may have interesting implications for various astrophysical processes, in particular structure/star/galaxy formation in the first billion years after the Big Bang (the so-called Cosmic Dawn). 
Here, we review from a theoretical standpoint how PBHs may impact the formation of the first stars and galaxies via their gravitational and non-gravitational effects.

Formed in the very early universe, PBHs are sources of gravity that produce additional perturbations in the matter density field, resulting in accelerated structure formation with respect to the standard $\Lambda$CDM case. In general, PBHs accelerate structure formation individually through the `seed' effect and collectively via the `Poisson' effect \cite{Carr:2018rid}. The `seed' effect dominates when the (mass) fraction of PBHs in DM is very small ($f_{\rm PBH}\rightarrow 0$), such that non-linear structures (i.e., DM halos) form around individual PBHs and do not interact with each other. In the idealized case where a single PBH with an initial mass $m_{\rm PBH}$ resides within a uniform background of particle DM (PDM), the mass of the halo seeded by the PBH follows a simple scaling law $M_{\rm B}\sim m_{\rm PBH}a/a_{\rm eq}$ in the matter-dominated epoch \cite{Mack:2006gz}, where $a=1/(1+z)$ is the current scale-factor and $a_{\rm eq}\sim 1/3400$ the scale factor at matter-radiation equality\footnote{This idealized picture of isolated growth of halos around PBHs will break down at late epochs when large-scale adiabatic perturbations have grown enough to drive clustering/interactions of PBH-seeded structures and/or when competition of accretion between individual PBH-seeded halos occurs at $a\gtrsim a_{\rm crit}\equiv a_{\rm eq}/f_{\rm PBH}$ with the critical scale factor $a_{\rm crit}$ defined by $M_{\rm B}(a_{\rm crit})= m_{\rm PBH}/f_{\rm PBH}$.}. 
The `Poisson' effect dominates in the opposite limit $f_{\rm PBH}\rightarrow 1$, as the random distribution of PBHs at small scales produce isocurvature Poisson perturbations in the primordial density field on top of the standard adiabatic/isentropic perturbations. The interplay between these two effects involves complex nonlinear dynamics in the intermediate case when PBHs do not make up all DM but are sufficiently abundant to allow interactions/mergers between DM halos induced by individual PBHs, which is still an active area of research (see, e.g., \cite{Inman:2019wvr,Liu:2022okz,Zhang:2024ytf} for the most recent progress with cosmological simulations). To zeroth order, the `seed' effect dominates at mass scales smaller than $\sim M_{\rm B}$, whereas the `Poisson' effect is important at mass scales around $\sim m_{\rm PBH}/f_{\rm PBH}$. 

In addition to the gravitational effects on DM structures that provide the gravitational potential wells for gas (baryonic matter) to condense and form stars/galaxies, PBHs also affect star/galaxy formation via the accretion of gas and the corresponding feedback. When gas falls onto a BH, part of its gravitational energy is converted into radiative, magnetic, and kinetic energy that can impact the surrounding medium through, e.g., ionization, heating, and shocks. This process of BH (accretion) feedback in general involves complex physics at multiple scales (from the Schwarzschild radius to a few Mpc), and it remains under vigorous theoretical and observational investigations (see, e.g., \cite{Abramowicz:2011xu,Silk:2013gqa,Yuan:2014gma,King:2015caa,Harrison:2017xcn,Fabbiano:2022njd,Fan:2022fhc}). Such `non-gravitational' effects should also be taken into account in a complete theoretical model of early star/galaxy formation under the influence of PBHs.


In this chapter, we introduce the key theoretical ingredients required to study the impact of PBHs on early star/galaxy formation and demonstrate their applications with examples of idealized PBH models and the relevant astrophysical context. The aim is to pave the way for future work that can further refine and extend these theoretical tools to consider more general PBH models and a broader range of astrophysical phenomena. 
For simplicity, we assume that all PBHs have the \textit{same} (initial) mass $m_{\rm PBH}$, and we focus on the stellar-mass ($\sim$10$ - 100\,\rm M_\odot$) and supermassive ($\sim$10$^7 - 10^{11}\,\rm M_\odot$) regimes motivated by recent GW observations and JWST results. We refer the reader to Part~II of the book for discussions on the PBH mass spectra from different formation mechanisms, some of which can indeed produce highly enhanced peaks like the ones assumed here (e.g., \cite{Carr:2018poi,Hooper:2023nnl}). 
We also ignore any additional clustering of PBHs beyond the small-scale random distribution at birth. Such clustering can occur in certain PBH formation mechanisms \cite{Belotsky:2018wph} (see Chapter~21), with non-trivial implications for the effects and observational signatures of PBHs \cite{Desjacques:2018wuu,Bringmann:2018mxj,DeLuca:2020jug} (see also Chapter 17 and 24). 

This chapter is organized as follows. In Section~\ref{sec:perturb}, we outline a heuristic theoretical framework for structure formation with PBHs, introducing the additional terms from PBHs in the linear power spectrum (Sec.~\ref{sec:powspec}) and a numerical scheme to generate initial conditions for cosmological simulations including PBHs (Sec.~\ref{sec:ic}). In Section~\ref{sec:accretion}, we describe a simple model for PBH accretion/feedback in terms of gas heating. In Section~\ref{sec:fsf}, based on the tools developed in previous sections, we combine semi-analytical analysis and cosmological simulations to investigate the effects of stellar-mass PBHs on primordial star formation, focusing on the cosmic stellar mass density (Sec.~\ref{sec:smd}), the internal structure of the first star-forming clouds/halos (Sec.~\ref{sec:halo}), and the formation of massive BH seeds (Sec.~\ref{sec:dcbh}). In Section~\ref{sec:smpbh}, we briefly discuss how the `seed' effect of SMPBHs could explain the unusually massive galaxy candidates observed by JWST with a back-of-the-envelope derivation. Finally, in Section~\ref{sec:dis}, we summarize our key findings and discuss the directions for future work. 

Throughout this chapter, unless otherwise noted, we adopt the cosmological parameters $\Omega_{m}=0.3089$, $\Omega_{b}=0.04864$, $\Omega_{r}=9.12\times 10^{-5}$, $n_{s}=0.9667$, $\sigma_{8}=0.8159$, and $h\equiv H_{0}/(100\ {\rm km\ s^{-1}\ Mpc^{-1}})=0.6774$ measured by the \textit{Planck} satellite for the standard $\Lambda$CDM cosmology \cite{Planck:2015fie}, which gives $a_{\rm eq}=\Omega_{r}/\Omega_{m}=2.952\times 10^{-4}$. 

\section{Structure formation within PBH cosmologies}
\label{sec:perturb}

In this section, we assemble a heuristic theoretical framework for cosmic structure formation in the presence of PBHs. 
For simplicity and in the absence of a better theory, we follow the formalism of linear perturbation theory developed for the standard $\Lambda$CDM case and introduce terms/treatments for the effects of PBHs. We first introduce the linear power spectrum of DM density fluctuations (extrapolated to the present time, $a=1$) that includes the `Poisson' effect from PBHs as additional isocurvature and correlation terms (Sec.~\ref{sec:powspec}). Then we describe a tentative numerical scheme to generate initial conditions for cosmological simulations with PBHs, designed to capture both `Poisson' and `seed' effects (Sec.~\ref{sec:ic}). 

\begin{warning}{Attention}
Due to the discrete and nonlinear nature of PBHs as point sources, linear perturbation theory is only strictly valid at large scales where we can treat PBHs as a quasi-homogeneous ideal pressure-less fluid like PDM. 
\end{warning}

\subsection{Power spectrum of DM density fluctuations with PBHs}
\label{sec:powspec}

Following the analysis in \cite{Inman:2019wvr}, we assume that both PDM and PBHs are subject to the same primordial adiabatic perturbations at all scales that cause the primordial overdensity $\delta^{0}_{\rm ad}$, while the primordial overdensity of PBHs has an additional term $\delta^{0}_{\rm iso}$ for isocurvature perturbations due to their discreteness that are important at small scales and assumed to be (initially) \textit{uncorrelated} with the adiabatic piece $\delta^{0}_{\rm ad}$, so the primordial overdensities of PDM and PBHs can be written as $\delta_{\rm PDM}(a\rightarrow 0)\equiv\delta^{0}_{\rm PDM}=\delta^{0}_{\rm ad}$ and $\delta_{\rm PBH}(a\rightarrow 0)\equiv\delta^{0}_{\rm PBH}=\delta^{0}_{\rm ad}+\delta^{0}_{\rm iso}$. 
The total DM overdensity is defined as $\delta_{\rm DM}\equiv (1-f_{\rm PBH})\delta_{\rm PDM}+f_{\rm PBH}\delta_{\rm PBH}$, whose evolution in the linear regime, starting from the initial condition $\delta^{0}_{\rm DM}=\delta^{0}_{\rm ad}+f_{\rm PBH}\delta^{0}_{\rm iso}$, follows 
\begin{align}
    \delta_{\rm DM}(a)=T_{\rm ad}(a)\delta^{0}_{\rm ad}+f_{\rm PBH}T_{\rm iso}(a)\delta^{0}_{\rm iso}\ .\label{delta_dm}
\end{align}
Here, $T_{\rm ad}(a)$ and $T_{\rm iso}(a)$ are the linear transfer functions for the adiabatic and isocurvature modes, respectively, which are normalized to unity at $a\rightarrow 0$. To derive the evolution of the sub-components $\delta_{\rm PDM}$ and $\delta_{\rm PBH}$, we further define the difference field $\delta_{-}\equiv \delta_{\rm PBH}-\delta_{\rm PDM}$. For simplicity, we assume that PBHs form with negligible velocities relative to PDM, such that $\delta_{-}=\delta_{-}^{0}=\delta^{0}_{\rm iso}$ remains constant according to the linearized continuity equations \cite{Inman:2019wvr}. In this way, we have
\begin{align}
    \delta_{\rm PDM}(a)&=\delta_{\rm DM}(a)-f_{\rm PBH}\delta^{0}_{\rm iso}=T_{\rm ad}(a)\delta^{0}_{\rm ad}+[T_{\rm iso}(a)-1]f_{\rm PBH}\delta^{0}_{\rm iso}\ ,\label{delta_pdm}\\
    \delta_{\rm PBH}(a)&=\delta_{\rm PDM}(a)+\delta^{0}_{\rm iso}=T_{\rm ad}(a)\delta_{\rm ad}^{0}+T_{\rm iso}^{\rm PBH}(a)\delta^{0}_{\rm iso}\ ,\label{delta_pbh}
\end{align}
where $T^{\rm PBH}_{\rm iso}(a)\equiv 1+\left[T_{\rm iso}(a)-1\right]f_{\rm PBH}$.

To appreciate the effects of PBHs on structure formation given the above formalism, let us focus on the total DM density fluctuations $\delta_{\rm DM}(a)$, whose statistical properties are well captured by the linear power spectrum $P_{\rm DM}(k,a)$, which in turn is the Fourier transform of the autocorrelation function. This can be written as
\begin{align}
    P_{\rm DM}(k,a)=T_{\rm ad}^{2}(a)P_{\rm ad}^{0}(k)+f_{\rm PBH}^{2}T_{\rm iso}^{2}(a)P_{\rm iso}^{0}(k)\ \label{pk}\ ,
\end{align}
ignoring any correlation between the adiabatic and isocurvature modes. 
Here, the first term is identical to the linear power spectrum in the standard $\Lambda$CDM case, which is well constrained at large ($k\lesssim 3h\ \rm Mpc^{-1}$) scales by observations (see fig.~19 in \cite{Planck:2018nkj}), while the second term captures the effects of PBHs. Since the PBH `seed' effect is by definition nonlinear, in this linear theory, we only consider the `Poisson' effect with
\begin{align}
P^{0}_{\rm iso}(k)\equiv 2\pi^{2}\Delta_{\rm Poisson}^{2}(k)/k^{3}=\bar{n}_{\rm PBH}^{-1}\ ,\quad k>k_{\rm eq}\ ,\label{poisson}
\end{align}
where $\Delta_{\rm Poisson}^{2}(k)=(k/k_{\star})^{3}$ is the variance of primordial Poisson perturbations given $k_{\star}=(2\pi^{2}\bar{n}_{\rm PBH})^{1/3}$ as the critical scale around the mean separation between PBHs and $\bar{n}_{\rm PBH}=f_{\rm PBH}[3H_{0}^{2}(\Omega_{m}-\Omega_{b})]/(8\pi Gm_{\rm PBH})$ as the cosmic average (co-moving) number density of PBHs. For simplicity, perturbations from PBHs are only included at scales smaller than the horizon scale at matter-radiation equality, defined by $k_{\rm eq}\sim 0.01\ \rm Mpc^{-1}$.

The linear power spectra can be used to calculate halo abundances semi-analytically using the Press-Schechter (PS) formalism \cite{Press:1973iz,mo2010galaxy}, which will be discussed in Sec.~\ref{sec:fsf} below. We note that in Eq.~\ref{pk}, we have ignored the correlation between isocurvature and adiabatic perturbations, so the halo abundances derived in this way should be regarded as \textit{conservative} estimates. Such correlation can be significant when PBHs fall into large structures due to adiabatic perturbations and meanwhile induce/disrupt PDM structures around themselves on small scales. 
Exploring this, cosmological simulations (see section~4 in \cite{Liu:2022okz}) have shown that the PS formalism based on the linear power spectrum without correlation (Eq.~\ref{pk}) indeed predicts less abundant halos than those in simulations in the halo mass range $0.5m_{\rm PBH}/f_{\rm PBH}\lesssim M_{\rm h}\lesssim 20m_{\rm PBH}/f_{\rm PBH}$. A heuristic term that captures the correlation can be included to achieve better agreements between the results of the PS formalism and simulations: 
\begin{align}
    P_{\rm DM}(k,a)&=T_{\rm ad}^{2}(a)P_{\rm ad}^{0}(k)+f_{\rm PBH}^{2}T_{\rm iso}^{2}(a)P_{\rm iso}^{0}(k)+P_{\rm corr}(k,a)\ ,\label{pk_new}\\
    P_{\rm corr}(k,a)&=f_{\rm PBH}T_{\rm iso}(a)\Delta_{\rm Poisson}^{2}(k)T_{\rm ad}^{2}(a)P_{\rm ad}^{0}(k),\quad k_{\rm eq}<k<3k_{\star}\ .\label{pk_corr}
\end{align}
Here the correlation term (Eq.~\ref{pk_corr}) is restricted to large scales ($k<3 k_{\star}$), because the small-scale behavior should be dominated by the Poisson noise in the PBH distribution. In fact, the effect of correlation is expected to be only important for $f_{\rm PBH}\lesssim 10^{-2}$ at intermediate scales where PBHs are clustered by adiabatic perturbations but still have significant Poisson noise in their distribution, and the gravity from PBHs do affect clustering of PDM but is not strong enough to seed or completely disrupt structures. 
More work needs to be done to validate this approach given the numerical difficulties in cosmological simulations (see the subsection below). Here we simply use the linear power spectrum with the heuristic correlation term (Eq.~\ref{pk_new}) to derive \textit{optimistic} estimates of the halo abundances under the effects of PBHs.

Considering the self-similar nature of perturbation growth in the linear regime, it is common to evaluate the (extrapolated) linear power spectrum at $a=1$. To do so, we use the \texttt{python} package \textsc{colossus} \cite{Diemer:2017bwl} to derive the adiabatic term $T^{2}_{\rm ad}(a=1)P^{0}_{\rm ad}(k)$ based on \textit{Planck} observations \cite{Planck:2015fie}, and calculate $T^{0}_{\rm iso}(a=1)$ with $T_{\rm iso}(a)=D(a)$ considering only the growing modes of isocurvature perturbations, where $D(a)$ is the linear growth rate given by the approximation formula from \cite{Inman:2019wvr}:
\begin{align}
    D(a)\approx \left(1+\frac{3\gamma a}{2a_{-}a_{\rm eq}}\right)^{a_{-}}\ ,\quad \gamma=\frac{\Omega_{m}-\Omega_{b}}{\Omega_{m}}\ ,\quad a_{-}=\frac{1}{4}\left(\sqrt{1+24\gamma}-1\right)\ .\label{dgrow}
\end{align}

\begin{figure}
\includegraphics[width=0.7\columnwidth]{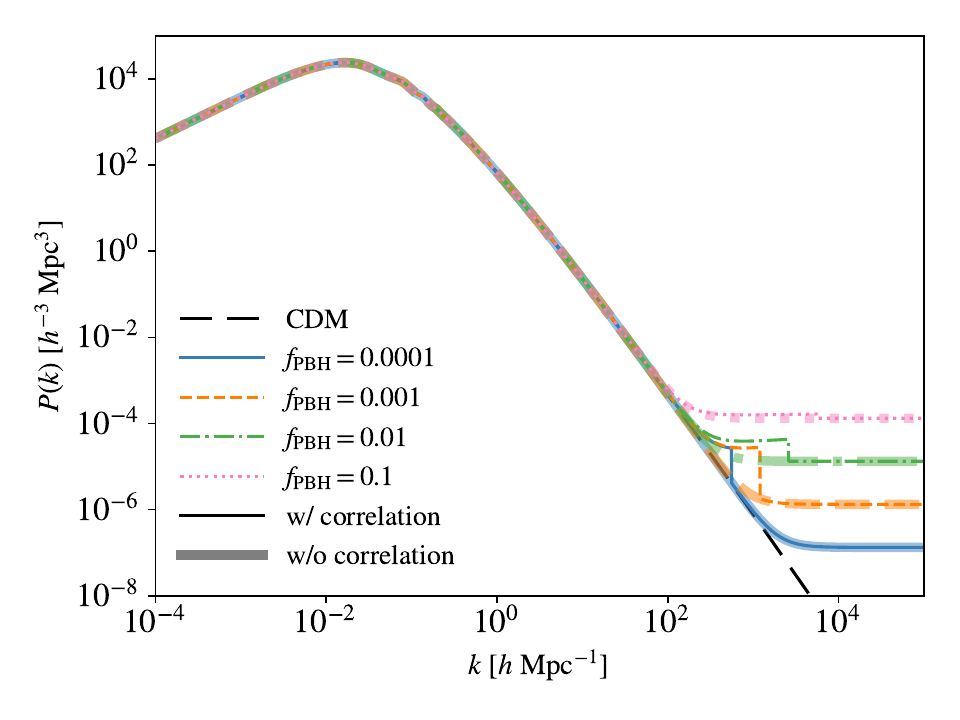}
\caption{Linear power spectra of DM density fluctuations (extrapolated to $a=1$) for the PBH models with $m_{\rm PBH}=33\ \rm M_{\odot}$, $f_{\rm PBH}=10^{-4}$ (solid), $10^{-3}$ (dashed), 0.01 (dashed-dotted), and 0.1 (dotted), in comparison with the reference power spectrum for the standard $\Lambda$CDM case (long-dashed) provided by the \texttt{python} package \textsc{colossus} \cite{Diemer:2017bwl} based on \textit{Planck} observations \cite{Planck:2015fie}, where two versions of the linear power spectrum with and without the heuristic correlation term (Eq.~\ref{pk_corr}) are shown for each PBH model with the thin and thick curves, respectively}
\label{fig:ps}       
\end{figure}

As examples, in Fig.~\ref{fig:ps} we show the linear power spectra at $a=1$ for the reference standard $\Lambda$CDM case ($f_{\rm PBH}=0$, long-dashed) and 4 PBH models with $m_{\rm PBH}=33\ \rm M_{\odot}$, $f_{\rm PBH}=10^{-4}$ (solid), $10^{-3}$ (dashed), 0.01 (dashed-dotted), and 0.1 (dotted). For each PBH model, two versions of the linear power spectrum with and without the heuristic correlation term are plotted with the thin and thick curves, respectively. 
It is evident that the Poisson noise term arising from PBHs overwhelms the intrinsic adiabatic term for large $k$, 
implying that structure formation will be accelerated by PBHs, especially at small scales and at early epochs. 



\subsection{Initial conditions for cosmological simulations with PBHs}
\label{sec:ic}
The linear power spectrum derived in the last subsection (Eqs.~\ref{pk}-\ref{pk_corr}) enables us to study the `Poisson' effect of PBHs on structure formation in a simple semi-analytical manner, as will be discussed in Sec.~\ref{sec:fsf}. However, this simple approach cannot capture the non-linear dynamics that shape the internal structure of and interactions between halos containing PBHs. One can model such nonlinear effects with cosmological (hydrodynamic) $N$-body simulations, in which gas can further be included in addition to DM to follow star and galaxy formation, as well as BH accretion and feedback (see, e.g., \cite{Sales:2022ich,Vogelsberger:2019ynw,Angulo:2021kes} for reviews). 

In order to utilize such advanced numerical machinery for our purposes, we must generate initial conditions for cosmological $N$-body simulations that include the effects of PBHs properly. In particular, we need to know the initial positions and velocities of PBHs and simulation particles/elements that embody (ensembles of) PDM. However, this is a challenging task, since state-of-the-art cosmological (hydrodynamic) simulation codes for star/galaxy formation (e.g., \textsc{gizmo} \cite{Hopkins:2014qka}, \textsc{arepo} \cite{Springel:2009aa} and \textsc{ramses} \cite{Teyssier:2001cp}) are designed to start in the matter-dominated epoch with initial redshifts of a few hundred when small-scale isocurvature perturbations of PBHs are already highly nonlinear while existing initial condition generators (e.g., \textsc{music} \cite{Hahn:2011uy}) only consider adiabatic perturbations that remain in the (quasi-)linear regime in this epoch. In this section, we outline a tentative numerical scheme for generating cosmological initial conditions with PBHs based on the exploratory work by \cite{Liu:2022okz}. We highlight the difficulties of modeling the nonlinear effects of PBHs in the initial conditions and discuss possible solutions that are yet to be improved in future studies. 

\subsubsection{Initial distribution of PBHs}\label{sec:pbh_dis}

We start with an initial simulation box at $a_{\rm ini}$ produced for the standard $\Lambda$CDM cosmology that only contains PDM (simulation) particles whose initial density and velocity fields reflect the standard adiabatic perturbations embodied by $T_{\rm ad}(a)\delta^{0}_{\rm ad}$ in Eqs.~\ref{delta_dm}-\ref{delta_pbh}. We then place PBHs into the box according to Eq.~\ref{delta_pbh} in two steps:
\begin{enumerate}
    \item We superimpose a {Cartesian} grid on the simulation box, whose cell size is chosen to have around one PBH in each cell on average, and calculate the local overdensity of DM in each cell $i$ as $\delta_{i}$. The cell size is meant to reflect the scale below which the discreteness of PBHs is important (see the next step).
    \item For each cell $i$, we draw the number of PBHs contained in it, $N_{i}$, from a Poisson distribution $p_{N}=\lambda_{i}^{N}e^{-\lambda_{i}}/(N!)$ with {$\lambda_{i}=(\delta_{i}+1)f_{\rm PBH}M_{\rm cell}/m_{\rm PBH}$, where $M_{\rm cell}$ is the average mass of DM per cell}. Then we place $N_{i}$ PBHs randomly in the cell. 
    \item Once the positions of PBHs are known, we assign a velocity to each PBH as the velocity of its nearest-neighbor PDM particle, assuming that PBHs form with negligible velocities relative to PDM. Finally, the mass of each PDM particle is reduced by a fraction of $f_{\rm PBH}$ for mass conservation.
\end{enumerate}
In this way, the fact that $\lambda_{i}$ is proportional to {$\delta_{i}$+1} and PBHs have the same velocity field as PDM captures the large-scale adiabatic perturbations on PBHs, i.e., the term $T_{\rm ad}(a)\delta^{0}_{\rm ad}$ in Eq.~\ref{delta_pbh}, while the random draw of $N_{i}$ and random spatial distribution of PBHs in each cell capture the isocurvature term $T_{\rm iso}^{\rm PBH}(a)\delta^{0}_{\rm iso}$. 

\subsubsection{Perturbations of PBHs on PDM and gas}\label{sec:pbh_pdm}

After addressing the PBH component, let us focus on the PDM component, for which adiabatic perturbations are already in place, and we only need to further implement the effects of PBHs on PDM denoted by the second term $[T_{\rm iso}(a)-1]f_{\rm PBH}\delta^{0}_{\rm iso}$ in Eq.~\ref{delta_pdm}. To do so, we first follow the Zel'dovich approximation \cite{Zeldovich:1969sb,mo2010galaxy} (which is commonly-used to produce initial conditions for the standard adiabatic perturbations) to associate the (co-moving) displacement $\vec{\psi}$ and (physical) velocity $\Delta\vec{v}$ fields produced by PBHs with their (co-moving) acceleration field $\nabla\phi_{\rm iso}(\vec{x})$:
\begin{align}
    &\vec{\psi}(\vec{x})=-\frac{2D_{\rm PBH}(a_{\rm ini})}{3\Omega_{m}H_{0}^{2}}\nabla \phi_{\rm iso}(\vec{x})\ ,\quad \Delta \vec{v}(\vec{x})=\frac{a_{\rm ini}\dot{D}_{\rm PBH}(a_{\rm ini})}{D_{\rm PBH}(a_{\rm ini})}\vec{\psi}(\vec{x})\ ,\label{zel}\\ 
    &\nabla \phi_{\rm iso}(\vec{x})=4\pi Gm_{\rm PBH}\sum_{i}\frac{\vec{x}-\vec{x}_{i}}{|\vec{x}-\vec{x}_{i}|^{3}}\ ,\label{acceleration}
\end{align}
given the (co-moving) coordinates of PBH particles\footnote{Since we are using the positions of PBHs at $a_{\rm ini}$ (rather than $a\rightarrow 0$) that are already affected by large-scale adiabatic perturbations, the {maximal correlation between adiabatic and isocurvature perturbations} is automatically taken into account here. } $\vec{x}_{i}$ at $a_{\rm ini}$. Here $D_{\rm PBH}(a_{\rm ini})$ is a growth factor, and we have $D_{\rm PBH}(a_{\rm ini})=T_{\rm iso}(a_{\rm ini})-1=D(a_{\rm ini})-1$ (considering only the growing modes) according to the linear theory (Eq.~\ref{delta_pdm}). Naively, with these equations, one can apply position and velocity changes to each PDM particle $j$ with original coordinate $\vec{x}_{j}$ and velocity $\vec{v}_{j}$ as $\vec{x}_{j}=\vec{x}_{j}+\vec{\psi}(\vec{x}_{j})$ and $\vec{v}_{j}=\vec{v}_{j}+\Delta\vec{v}(\vec{x}_{j})$.
This simple approach is only strictly valid in the linear regime when $|\vec{\psi}|$ is small, but perturbations to particles very close to PBHs are nonlinear (i.e., the `seed' effect). Considering a single PBH for simplicity, a natural definition of the nonlinear scale is $d_{\rm nl}=[D_{\rm PBH}(a_{\rm ini})/\bar{n}_{\rm PBH}]^{1/3}$, since when the distance between a PDM particle to the PBH $r$ is below $d_{\rm nl}$, we will have $|\vec{\psi}|\gtrsim r$ according to Eqs.~\ref{zel} and \ref{acceleration}. If we start the simulation deep in the radiation-dominated regime ($a_{\rm ini}\ll a_{\rm eq}$) when $D_{\rm PBH}\approx 1.5\gamma a_{\rm ini}/a_{\rm eq}$ (see Eq.~\ref{dgrow}) is very small, this is not a serious problem as $d_{\rm nl}$ will usually be much smaller than the (initial) resolution of the simulation.

However, to study star/galaxy formation with existing simulation codes, we want to start the simulation in the matter-dominated epoch with $a_{\rm ini}\sim 0.001-0.01$, because it is computationally intensive to simulate baryonic physics at very high redshifts. In this case, the exact form of $D_{\rm PBH}(a_{\rm ini})$ can be different from what is expected/extrapolated from the linear theory, and $d_{\rm nl}$ can be larger than the simulation resolution or even larger than the structures that we are interested in. Here, we introduce two numerical treatments to address these issues targeting the `Poisson' and `seed' effects, respectively, which will be combined in the end to construct a tentative scheme that captures both effects.

First, if we are only concerned with the formation of structures much larger than $d_{\rm nl}$ at late epochs under the collective influence of multiple PBHs (i.e., the `Poisson' effect), we can insure linearity everywhere in the initial setup, by applying a truncation on the displacement with \cite{Inman:2019wvr,Liu:2022okz}
\begin{align}
    \vec{\psi}=\min\left(1,d_{\rm PDM}/|\vec{\psi}|\right)\vec{\psi}\ ,\label{trunc_grid}
\end{align}
given $d_{\rm PDM}$ as the average distance between PDM particles, which is a measure of the (initial) resolution of the simulation. Here, the truncation will only delay structure formation around PBHs at small scales (i.e., the `seed' effect), while the perturbations from PBHs that affect large structures are still small in the initial condition and are therefore unaffected by the truncation. 

Next, to capture the nonlinear structures around individual PBHs (i.e., the `seed' effect), we often need stronger perturbations than allowed by the truncation (Eq.~\ref{trunc_grid}) in the region very close to a PBH (i.e., within $d_{\rm nl}$). Considering a PDM particle perturbed by a single PBH at a distance $r$ away from it, the displacement $\vec{\psi}$ points from the PDM particle to the BH, and we have $|\vec{\psi}|\gtrsim r$ for $r\lesssim d_{\rm nl}$, which indicates that nonlinear dynamics (e.g., shell crossing) is already important in this region. The final outcome of such nonlinear dynamics is a virialized halo with a mass of $M_{\rm B}\sim m_{\rm PBH}a/a_{\rm eq}$ \cite{Mack:2006gz} and an overdensity of $\Delta_{\rm vir}=18\pi^{2}\sim 200$ according to the spherical collapse theory (see Chapter 5 in \cite{mo2010galaxy}), which is also what we want to produce in an idealized simulation of one single PBH. In light of this, we can apply the following truncation 
\begin{align}
    \vec{\psi}=\min\left(1,f_{\rm shrink}r/|\vec{\psi}|\right)\vec{\psi}\ ,\quad f_{\rm shrink}=1-\Delta_{\rm vir}^{-1/3}\approx 0.822\ ,\label{trunc_seed}
\end{align}
to create a uniform sphere of PDM around the PBH with an overdensity of $\Delta_{\rm vir}$ and a radius of $\sim d_{\rm nl}$, mimicking the non-linear structure seeded by the PBH (see \cite{Jiao:2024rcr} for another approach to creating nonlinear structures around point-sources of gravity in the initial condition based on virialization). We find by test simulations 
of a single PBH starting from $a_{\rm ini}\in [a_{\rm eq},0.01]$ with the codes \textsc{gizmo} \cite{Hopkins:2014qka} and \textsc{arepo} \cite{Springel:2009aa} (whose gravity solvers are both based on \textsc{gadget-2} \cite{Springel:2005mi}) that the initial uniform sphere will quickly collapse and virialize within a local Hubble time to become a halo\footnote{We use the code \textsc{rockstar} \cite{Behroozi:2011ju} to identify halos in our simulations and calculate halo (virial) masses and radii.} whose mass grows linearly with $a$ as $M_{\rm h}\sim (a/a_{\rm ini})(4\pi/3)D_{\rm PBH}(a_{\rm ini})m_{\rm PBH}$. This linear relation with $a$ is consistent with $M_{\rm B}\sim m_{\rm PBH}a/a_{\rm eq}$ predicted by the spherical collapse theory \cite{Mack:2006gz}, but the normalization is off by a factor of a few if we derive the growth factor from the linear growth rate $D_{\rm PBH}(a_{\rm ini})=D(a_{\rm ini})-1$ (see Eq.~\ref{dgrow}). This implies that the strength of perturbations is overestimated by the (extrapolated) linear theory or there are numerical issues in our simulations. In fact, even if we do not consider the nonlinear structure initially and use the stronger truncation in Eq.~\ref{trunc_grid}, the halo mass will converge to $M_{\rm h}\sim (a/a_{\rm ini})(4\pi/3)D_{\rm PBH}(a_{\rm ini})m_{\rm PBH}$ at late epochs. In light of this, we adopt $D_{\rm PBH}(a_{\rm ini})\sim 3a_{\rm ini}/(4\pi a_{\rm eq})$ to ensure $M_{\rm h}\sim M_{\rm B}$. As $D_{\rm PBH}(a)\propto a$, we have $\dot{D}_{\rm PBH}(a)/D_{\rm PBH}(a)=\dot{a}/a=H(a)$, which can be used to derive $\Delta\vec{v}$ (Eq.~\ref{zel}).

Finally, we combine the two prescriptions by dividing the perturbation of PBHs on a PDM particle into two pieces $\vec{\psi}=\vec{\psi}_{\rm seed}+\vec{\psi}_{\rm Poisson}$. Here $\vec{\psi}_{\rm seed}$ arises from the `seed' effect of the nearest PBH, while $\vec{\psi}_{\rm Poisson}$ captures the collective `Poisson' effect of all the other PBHs. In practice, we only consider the nearest PBH in Eq.~\ref{acceleration} to calculate $\vec{\psi}_{\rm seed}$, which is subject to the truncation based on halo overdensity (Eq.~\ref{trunc_seed}). For $\vec{\psi}_{\rm Poisson}$, we add up the accelerations from all the other PBHs in Eq.~\ref{acceleration} and use the truncation based on resolution (Eq.~\ref{trunc_grid}). 
When gas particles are included in the initial condition, we apply PBH perturbations to them following the same procedure developed for PDM (Eq.~\ref{zel}-\ref{trunc_seed}) but with a smaller growth factor $D_{\rm PBH}(a_{\rm ini})\sim 3a_{\rm ini}/(4\pi a_{\rm rec})$, assuming that isocurvature perturbations in gas only starts to grow after photon-gas decoupling (i.e., recombination) at $a_{\rm rec}=1/(1+z_{\rm rec})\simeq 1/1100$. 

We find by dark-matter-only test simulations with \textsc{gizmo} \cite{Hopkins:2014qka} for $f_{\rm PBH}\le 0.1$ that the relation $M_{\rm h}\sim M_{\rm B}\sim m_{\rm PBH}a/a_{\rm eq}$ holds at early stages dominated by the `seed' effect, while the mass distributions of halos from simulations are consistent with the predictions from the PS formalism\footnote{However, the abundance of smaller ($M_{\rm h}\lesssim 0.5m_{\rm PBH}/f_{\rm PBH}$) halos in PBH simulations is significantly lower than that predicted by the PS formalism (by up to two orders of magnitude), even lower than the abundance in the standard $\Lambda$CDM case. The reason is that the formation and growth of small halos without PBHs 
are suppressed by the disruption and engulfment of nearby PBH-induced halos \cite{Zhang:2024ytf}. The PS formalism also overproduces large ($M_{\rm h}\gtrsim 20m_{\rm PBH}/f_{\rm PBH}$) halos by up to a factor of 10 for $f_{\rm PBH}\gtrsim 0.01$. The assembly of such massive halos containing many PBHs are likely delayed by the tightly bound (sub)structures seeded by PBHs. This is consistent with the finding in \cite{Inman:2019wvr} that the non-linear growth of large-scale fluctuations caused by PBH is slower than expected from the linear theory extrapolation. 
These discrepancies (see fig.~14 in \cite{Liu:2022okz}) indicate that the PS formalism cannot fully capture the non-linear dynamics and non-Gaussianity caused by PBHs in structure formation.} based on the linear power spectrum (Eq.~\ref{pk_new}) including the correlation term (Eq.~\ref{pk_corr}) at late epochs dominated by the `Poisson' effect for haloes with {$0.5m_{\rm PBH}/f_{\rm PBH}\lesssim M_{\rm h}\lesssim 20m_{\rm PBH}/f_{\rm PBH}$.} 






%





\section{PBH accretion and feedback}
\label{sec:accretion}

Beyond structure formation, the evolution of gas and radiation at Cosmic Dawn is also regulated by PBHs via BH accretion and feedback. Current studies focus on the radiation and heating produced by accreting PBHs. The former contributes to a variety of cosmic radiation backgrounds (\cite{Ricotti:2007au,Hasinger:2020ptw,Cappelluti:2021usg,Mittal:2021egv,Acharya:2022txp,Ziparo:2022fnc,Zhang:2023hyn,Manzoni:2023gon}) and affects the thermal and ionization histories of the intergalactic medium (IGM), which provides important evidence/constraints for PBHs (see Part V of this book). The latter alters the thermal and chemical properties of the interstellar medium (ISM, \cite{Lu:2020bmd,Laha:2020vhg,Takhistov:2021aqx,Casanueva-Villarreal:2024ifm}) that are crucial for star formation. In this section, we describe a simple model to calculate the heating rate from an accreting PBH as a function of environmental parameters, which will be used in Sec.~\ref{sec:fsf} to evaluate the effects of PBHs on the first star-forming halos. The reader is referred to Chapter 14 {(see also Chapter 26)} for a broader and more detailed discussion on PBH accretion and feedback, in particular their impact on the evolution of PBHs themselves. 

The mass accretion rate $\dot{m}_{\mathrm{acc}}$ of a BH with a mass of $m_{\rm BH}$ surrounded by a medium with a gas density of $\rho_{\rm gas}$ is given by the Bondi–Hoyle–Lyttleton formula
\begin{align}
    \dot{m}_{\mathrm{acc}}=4\pi r_{\rm B}^{2}\tilde{v}\rho_{\rm gas}=\frac{4\pi (G m_{\mathrm{BH}})^{2}\rho_{\mathrm{\rm gas}}}{\tilde{v}^{3}}\ ,\label{bondi}
\end{align}
where $r_{\rm B}=Gm_{\rm BH}/\tilde{v}^{2}$ is the Bondi radius, and $\tilde{v}=\sqrt{v_{\rm rel}^{2}+c_{s}^{2}}$ is the characteristic velocity. 
Here $v_{\rm rel}$ is the relative velocity between the BH and the medium, and $c_{s}$ is the gas sound speed. 

The bolometric luminosity $L_{\rm BH}$ of the accretion flow is then related to the accretion rate by $L_{\rm BH}=\epsilon_{\rm EM}\dot{m}_{\rm acc}c^{2}$, where $\epsilon_{\rm EM}$ is the radiation efficiency which depends on the detailed properties of the accretion flow and to the first order is a function of accretion rate and BH mass. For simplicity, we use the formula of $\epsilon_{\rm EM}$ in \cite{Negri:2016epd} to capture the transition from optically thick and geometrically thin, radiatively efficient accretion disks, to optically thin, geometrically thick, radiatively inefficient advection dominated accretion flows:
\begin{align}
\epsilon_{\mathrm{EM}}=\frac{\epsilon_{0}A\eta}{1+A\eta}\ ,\quad \eta\equiv \dot{m}_{\mathrm{acc}}/\dot{m}_{\mathrm{Edd}}\ ,\label{epsilonEM}
\end{align}
where $A=100$, $\epsilon_{0}=0.057$ is the radiative efficiency for non-rotating Schwarzschild BHs (assuming negligible spins of PBHs for simplicity), and $\eta$ is the Eddington ratio defined by the Eddington accretion rate $\dot{m}_{\mathrm{Edd}}$, which for primordial gas made of $\sim 75\%$ hydrogen and $\sim 25\%$ helium (mass fractions) is 
\begin{align}
	\dot{m}_{\mathrm{Edd}}\simeq 2.7\times 10^{-6}\ \mathrm{M_{\odot}\ yr^{-1}}\ \left(\frac{m_{\mathrm{BH}}}{100\ \mathrm{M_{\odot}}}\right)\left(\frac{\epsilon_{0}}{0.1}\right)^{-1}\ .\label{medd}
\end{align}
It is shown in \cite{Liu:2022okz} that Eq.~\ref{epsilonEM} is consistent with the results of detailed calculations of the spectra of BH accretion flows based on \cite{Takhistov:2021aqx}.

Now, given Eqs.~\ref{bondi}-\ref{medd}, 
we can derive the heating rate of an accreting BH as
\begin{align}
    P(m_{\rm BH},\rho_{\rm gas},\tilde{v})=f_{\rm abs}f_{h}\epsilon_{\rm EM}(m_{\rm BH},\dot{m}_{\rm acc})\dot{m}_{\rm acc}(m_{\rm BH},\rho_{\rm gas},\tilde{v})c^{2}\ ,\label{P_heat}
\end{align}
where $f_{\rm abs}$ is the fraction of radiation energy (carried by ionizing photons) absorbed by the system that we are concerned with (e.g., primordial star-forming halos), and $f_{h}$ is the fraction of energy deposited into the ISM as heat. We assume a constant absorption faction $f_{\rm abs}= 0.66$ for simplicity 
according to radiative transfer calculations for typical primordial star-forming clouds in \cite{Liu:2022okz}, and adopt $f_{h}=1/3$ following \cite{Lu:2020bmd,Takhistov:2021aqx}.


\section{Primordial star formation under the influence of stellar-mass PBHs}
\label{sec:fsf}

In this section, we explore the effects of stellar-mass PBHs on primordial star formation. We focus on the particular case with an initial PBH mass of $m_{\rm PBH}=33\ \rm M_\odot$. We choose this mass because it is the location of the (secondary) Gaussian peak in the best-fit Power-Law + Peak model of the mass distribution of BHs detected in GW observations of binary BH mergers by the LIGO-Virgo-KAGRA Scientific Collaboration \cite{LIGOScientific:2020kqk}. Varying $m_{\rm PBH}$ by a factor of a few does not change our conclusions. We focus on PBH (initial) mass fractions in the range $f_{\rm PBH}\in [10^{-4}-0.1]$ with $f_{\rm PBH}=10^{-3}$ being the fiducial case, motivated by existing constraints (and their uncertainties) on PBH abundance \cite{Carr:2020gox} (see also Part V of the book). Here $f_{\rm PBH}\gtrsim 0.01$ is strongly disfavored for $m_{\rm PBH}\sim 30\ \rm M_\odot$ by observations from GWs, the 21-cm signal and CMB. We still consider the extreme regime of $f_{\rm PBH}\sim 0.01-0.1$ for the sake of theoretical exploration. 

In the standard $\Lambda$CDM cosmology, the first generation of stars, the so-called Population~III (Pop~III), form typically in minihalos of $M_{\rm h}\sim 10^{5}-10^{8}\ \rm M_{\odot}$ at $z\sim 10-30$ where \textit{primordial} (pure H/He) gas clouds collapse via cooling by molecular hydrogen $\rm H_{2}$ in the absence of metals (see, e.g., \cite{Bromm:2013,Klessen:2023qmc} for reviews). Here the minimum halo mass required to form Pop~III stars is $M_{\rm mol}\sim 10^{5}-10^{7}\ \rm M_\odot$ \cite{Trenti:2009cj,Schauer:2020gvx}, which is determined by the criterion of efficient molecular cooling and is a function of redshift and environmental parameters (see below). The upper bound reflects the fact that massive halos ($M_{\rm h}\gtrsim 10^{8}\ \rm M_\odot$) tend to be metal-enriched and form second-generation stars \cite{Bromm:2011}. The abundance of such massive halos is also much lower than that of low-mass Pop~III star-forming halos at high $z$. Therefore, we focus on $M_{\rm h}<10^{8}\ \rm M_\odot$ when considering Pop~III star formation. 

In our case with $m_{\rm PBH}=33\ \rm M_\odot$, the abundance of halos that host Pop~III stars in the standard $\Lambda$CDM cosmology is enhanced by PBHs due to their `Poisson' effect (Sec.~\ref{sec:powspec}), while the `seed' effect is expected to be unimportant because the mass of an isolated halo seeded by a PBH is $M_{\rm B}\lesssim 10^{4}\ {\rm M_\odot}$ at $z\gtrsim 10$, much smaller than $M_{\rm mol}$. On the other hand, PBHs accrete and heat the surrounding gas, such that stronger (molecular) cooling is required for the clouds to collapse and form stars, leading to higher $M_{\rm mol}$, which tends to reduce the abundance of Pop~III star-forming halos. Here we first consider these two competing effects self-consistently in a semi-analytical analysis to evaluate the global impact of PBHs on the cosmic mass density of Pop~III stars ever formed in the universe (Sec.~\ref{sec:smd}). Then we look into the local effects of PBHs on the internal structure of Pop~III star-forming halos and clouds using the results from the cosmological zoom-in simulations in \cite{Liu:2022okz} (Sec.~\ref{sec:halo}). Finally, we briefly discuss the potential impact of PBHs on the formation of massive BH seeds in more massive halos ($M_{\rm h}\gtrsim 10^{8}\ \rm M_{\odot}$) in Sec.~\ref{sec:dcbh}.

\subsection{Cosmic stellar mass density of Pop~III stars}\label{sec:smd}

\begin{figure}
\includegraphics[width=1\columnwidth]{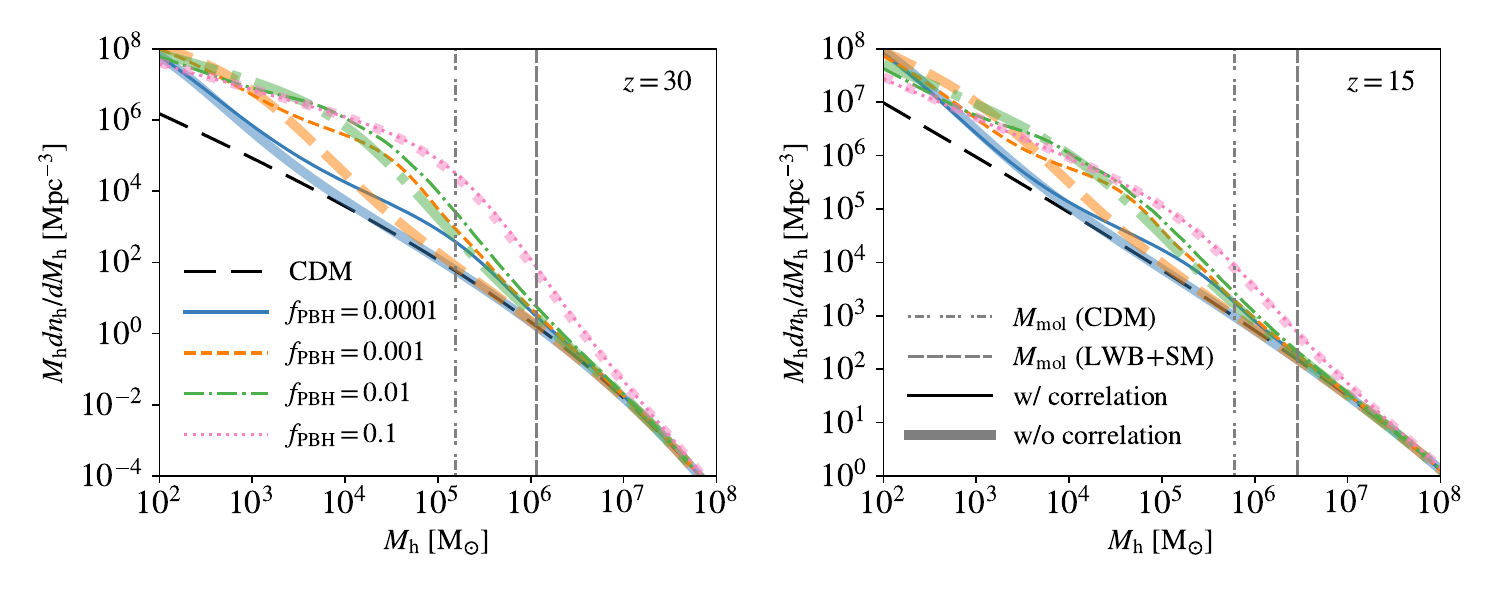}
\caption{Halo mass functions for the PBH models with $m_{\rm PBH}=33\ \rm M_{\odot}$, $f_{\rm PBH}=10^{-4}$ (solid), $10^{-3}$ (dashed), 0.01 (dashed-dotted), and 0.1 (dotted), compared with the $\Lambda$CDM reference mass function (long-dashed), at $z=30$ (\textit{left}) and $z=15$ (\textit{right}), where two versions of the halo mass function with and without the heuristic correlation term in the linear power spectrum (Eq.~\ref{pk_corr}) are shown for each PBH model with the thin and thick curves, 
and the dot-dash-dotted and densely dashed vertical lines label the mass thresholds of Pop~III star formation via molecular cooling in the standard $\Lambda$CDM case with and without the environmental effects of LWB and SM \cite{Trenti:2009cj,Schauer:2020gvx}, respectively}
\label{fig:hmf}       
\end{figure}

To appreciate the acceleration of structure formation by the `Poisson' effect of PBHs, we calculate the halo mass function $dn_{\rm h}/dM_{\rm h}$ (co-moving number density of halos per unit halo mass) with the PS formalism (see Chapter 7 of \cite{mo2010galaxy} for details) using the linear power spectrum with enhanced small-scale power by PBHs (Eqs.~\ref{pk}-\ref{pk_corr}). We adopt the commonly-used top-hat window function to derive the density variance at a given mass scale (see Sec. 6.1.3 in \cite{mo2010galaxy}) and consider the correction for ellipsoidal dynamics (Eq.~7.67 in \cite{mo2010galaxy}) that is required for good agreements between the predictions of the PS formalism and $N$-body simulations. The results are shown in Fig.~\ref{fig:hmf} for $f_{\rm PBH}=10^{-4}$ (solid), $10^{-3}$ (dashed), 0.01 (dashed-dotted), and 0.1 (dotted), compared with the standard $\Lambda$CDM results (long-dashed) at $z=30$ (left) and $z=15$ (right). Similar to Fig.~\ref{fig:ps}, the halo mass functions for two versions of the linear power spectrum with and without the heuristic correlation term (Eq.~\ref{pk_corr}) are plotted with the thin and thick curves (as optimistic and conservative estimates) for each PBH model, respectively. Clearly, compared with the standard $\Lambda$CDM case (long-dashed), the abundances of halos are boosted by PBHs by up to $\sim 3$ orders of magnitude for $M_{\rm h}\sim 10^{5}\ \rm M_\odot$ at $z=30$ in the extreme model with $f_{\rm PBH}=0.1$. The effect is stronger at higher redshift. Taking into account the correlation term (Eq.~\ref{pk_corr}) in the linear power spectrum (thick curves) significantly increases the halo abundance at $M_{\rm h}\gtrsim M_{\rm B}\sim$ a few thousand $\rm M_\odot$ for $f_{\rm PBH}\lesssim 0.01$. 

\begin{warning}{Attention}

The Press-Schechter formalism assumes that density fluctuations are Gaussian. However, in the presence of PBHs, the density field is non-Gaussian, especially at small scales dominated by isocurvature perturbations from the Poisson distributed PBHs \cite{Chisholm:2005vm,Desjacques:2018wuu,Inman:2019wvr}. 
Our calculation of the halo mass function with the Press-Schechter formalism here is an approximation commonly used in previous studies \cite{Kashlinsky:2016,Kashlinsky:2021,Cappelluti:2021usg,Atrio-Barandela:2022jov}. Cosmological simulations with proper initial conditions are required to accurately derive the halo mass functions in PBH cosmologies (see Sec.~\ref{sec:ic}). 
\end{warning}

To interpret these results in the context of Pop~III star formation, in Fig.~\ref{fig:hmf} we also plot the halo mass thresholds of Pop~III star formation $M_{\rm mol}$ in the standard $\Lambda$CDM cosmology for 2 situations: First (dot-dash-dotted), in the idealized case without any environmental effects (e.g., halo assembly history, metal enrichment, radiative feedback, cosmic rays, and streaming motion between DM and gas), we have 
\begin{align}
    M_{\rm mol}\sim 1.54\times 10^{5}\ {\rm M_\odot}[(1+z)/31]^{-2.074}\label{mmol}
\end{align}
according to the simple analysis of primordial chemistry and cooling in \cite{Trenti:2009cj}. Second (densely-dashed), we consider the environmental effects of Lyman-Werner (LW) radiation and streaming motion\footnote{With PBHs, the effects of gas-DM streaming are weaker than those in the standard $\Lambda$CDM case due to structures induced by PBHs, which can facilitate gas collapse in star-forming halos for $f_{\rm PBH}m_{\rm PBH}\gtrsim 3\ \rm M_{\odot}$ \cite{Kashlinsky:2021}. Here, we ignore such effects for simplicity, as we are mostly concerned with PBH models with $f_{\rm PBH}m_{\rm PBH}\lesssim 3\ \rm M_{\odot}$, although similar trends are also observed in cosmological zoom-in simulations for $f_{\rm PBH}=0.001$ and $m_{\rm PBH}=33\ \rm M_\odot$ \cite{Liu:2022okz} (see also \cite{Atrio-Barandela:2022jov}).} (SM) between DM and baryons with the fit formula of $M_{\rm mol}(J_{21},v_{\rm bc})$ derived from the cosmological simulations in \cite{Schauer:2020gvx} (see their Eqs.~11-13):
\begin{align}
\begin{split}
    &\log(M_{\rm mol}/{\rm M_\odot})=\log(M_{0}/{\rm M_\odot})+b\times v_{\rm bc}/\sigma_{\rm rms}\ ,\\
    &\log(M_{0}/{\rm M_\odot})=5.562\times(1+0.279\sqrt{J_{21}})\ ,\quad b=0.614\times (1-0.56\sqrt{J_{21}})\ ,
\end{split}\label{mmol_ext}
\end{align}
where $J_{21}$ is the intensity of LW background (LWB) as a function of redshift in units of $10^{-21}\ \rm erg\ s^{-1}\ cm^{-2}\ Hz^{-1}\ sr^{-2}$, and $v_{\rm bc}/\sigma_{\rm rms}$ is the relative velocity between DM and gas in units of the SM root-mean-square velocity $\sigma_{\rm rms}$. We adopt $J_{21}=10^{2-z/5}$ from \cite{Greif:2006nr} and $v_{\rm bc}/\sigma_{\rm rms}=0.8$ as the most common SM velocity \cite{Schauer:2020gvx} to capture the typical levels of LWB and SM. Such environmental effects increase $M_{\rm mol}$ by a factor of $\sim 8\ (5)$ at $z=30\ (15)$ compared with the idealized case (Eq.~\ref{mmol}) because the LWB dissociates $\rm H_{2}$ and the SM engenders additional turbulence energy in the primordial cloud, thus delaying collapse. We find that for $f_{\rm PBH}\sim 10^{-4}-0.01$ (allowed by observational constraints) the abundance of halos above $M_{\rm mol}$ is significantly enhanced only in the idealized case (Eq.~\ref{mmol}), and only when the correlation term is considered for $f_{\rm PBH}\sim 10^{-4}-10^{-3}$. The effects of PBHs are much weaker for more massive halos above the mass threshold subject to the environmental effects of LWB and SM (Eq.~\ref{mmol_ext}), such that only the extreme model with $f_{\rm PBH}=0.1$ exhibits significant differences.

To quantify the effects of PBHs, let us consider the impact of PBH accretion heating on the mass threshold of Pop~III star formation. For simplicity, we focus on the idealized case without environmental effects. To the first order, the mass threshold $M_{\rm mol}$ corresponds to the minimum halo mass that satisfies the Rees-Ostriker-Silk cooling criterion \cite{Rees1977,Silk:1977wz} $t_{\mathrm{cool}}\le t_{\mathrm{ff}}$, where $t_{\rm cool}$ is the cooling timescale, and $t_{\rm ff}$ is the free-fall timescale:
\begin{align}
    t_{\rm cool}&=\frac{(3/2)k_{B}T_{\rm vir}}{\Lambda(T_{\rm vir},n_{\rm H})x_{\rm H_{2}} - \Gamma (m_{\rm PBH}, f_{\rm PBH}, \rho_{\rm gas}, \tilde{v})}\ ,\\
    t_{\rm ff}&=\sqrt{\frac{3\pi}{32G\rho_{\rm gas}}}\ .
\end{align} 
Here $T_{\rm vir}$ is the virial temperature of the halo (see Eq.~2 in \cite{Trenti:2009cj}), $x_{\rm H_{2}}$ is the maximum abundance of $\rm H_{2}$ (i.e., ratio of number densities between $\rm H_{\rm 2}$ and hydrogen nuclei), {$\rho_{\rm gas}=X^{-1}m_{\rm H}n_{\rm H}=\Delta_{\rm gas}\bar{\rho}_{\rm gas}$} is the characteristic (physical) density of gas defined by the overdensity parameter $\Delta$ and the cosmic average gas density $\bar{\rho}_{\rm gas}$, $n_{\rm H}$ is the number density of hydrogen nuclei related to $\rho_{\rm gas}$ by the primordial hydrogen mass fraction $X\simeq 0.76$ and proton mass $m_{\rm H}$, $\Lambda$ is the cooling rate per $\rm H_{2}$, and $\Gamma$ is the heating rate per baryon from PBH accretion. 

To evaluate the cooling rate, we adopt the approximation formulae of $x_{\rm H_{2}}$ and $\Lambda$ in \cite{Trenti:2009cj} for the standard $\Lambda$CDM case, and further consider the enhancement of $\rm H_{2}$ formation due to ionization of gas by accreting PBHs found in cosmological simulations \cite{Liu:2022okz} (captured by the multiplication term with $f_{\rm PBH}$ in Eq.~\ref{xh2}):
\begin{align}
    &x_{\rm H2}\simeq 3.5\times 10^{-4}\left(\frac{T_{\rm vir}}{1000\ \rm K}\right)^{1.52}\times \max[1,3(f_{\rm PBH}/0.1)^{0.15}]\ ,\label{xh2}\\
    &\Lambda(T_{\rm vir},n_{\rm H})\simeq 10^{-31.6}\ {\rm erg\ s^{-1}}\left(\frac{T}{100\ {\rm K}}\right)^{3.4}\left(\frac{n_{\rm H}}{10^{-4}\ \rm cm^{-3}}\right)\ .\label{lambda}
\end{align}
The heating rate per baryon is related to the heating rate per BH (Eq.~\ref{P_heat}) by
\begin{align}
    \Gamma=\frac{f_{\rm PBH}\mu m_{\rm H}(\Omega_{m}-\Omega_{b})}{m_{\rm PBH}\Omega_{b}}P(m_{\rm PBH}, \rho_{\rm gas}, \tilde{v})\ ,
\end{align}
where {$\mu\simeq 1.22$ is the mean molecular weight of primordial gas, and} we estimate the characteristic velocity between PBHs and gas as
\begin{align}
    \tilde{v}\sim \sqrt{\frac{GM_{\rm h}}{R_{\rm vir}}}\sim {5.4\ \rm km\ s^{-1}} \left(\frac{M_{\rm h}}{10^{6}\ \rm M_{\odot}}\right)^{\frac{1}{3}}\left(\frac{21}{1+z}\right)^{\frac{1}{2}}\ .\label{e22}
\end{align}
Here for simplicity, we use $m_{\rm BH}=m_{\rm PBH}$ to evaluate the heating rate per BH (Eq.~\ref{P_heat}) considering that mass growth by accretion is minor ($\lesssim 10\%$) at $z\gtrsim 10$ for our stellar-mass PBHs with an initial mass of $m_{\rm PBH}=33\ \rm M_\odot$, as shown in \cite{Liu:2022okz} (see also Chapter 14 {and \cite{Gordon:2024odc}}). Finally, we set the overdensity parameter as $\Delta_{\rm gas}=1300$ to reproduce the results from \cite{Trenti:2009cj} for the $\Lambda$CDM case (Eq.~\ref{mmol}) with $f_{\rm PBH}=0$. 

\begin{figure}
\sidecaption
\includegraphics[width=0.7\columnwidth]{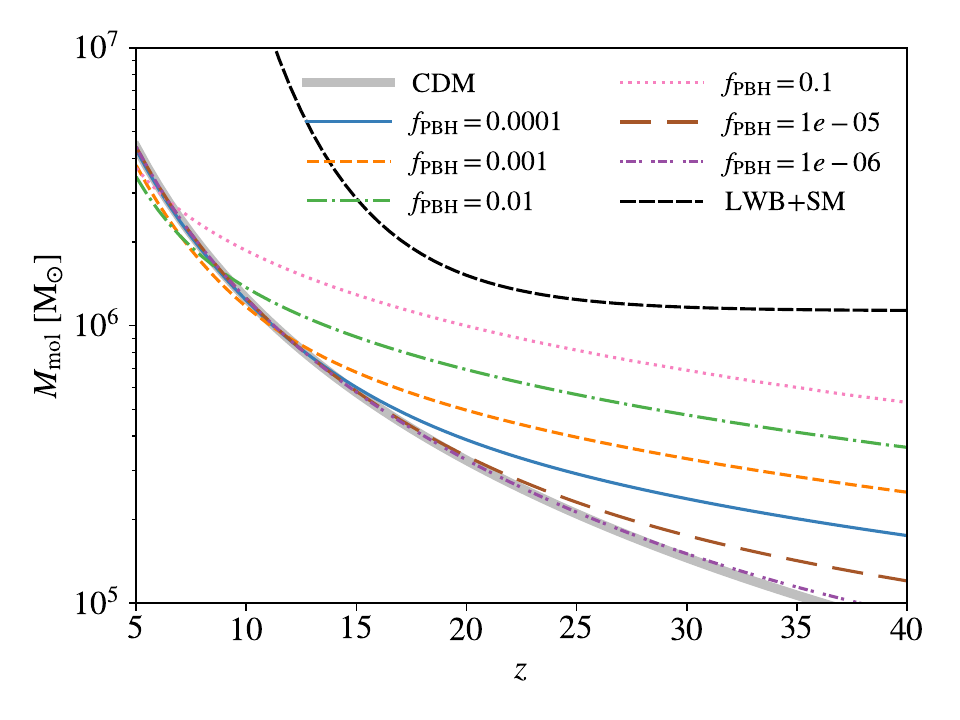}
\caption{Halo mass thresholds of Pop~III star formation via molecular cooling as functions of redshift without environmental effects, for the PBH models with $m_{\rm PBH}=33\ \rm M_{\odot}$, $f_{\rm PBH}=10^{-4}$ (solid), $10^{-3}$ (dashed), 0.01 (dashed-dotted), 0.1 (dotted), $10^{-5}$ (long-dashed), and $10^{-6}$ (dot-dash-dotted), compared with the standard $\Lambda$CDM case (thick solid, Eq.~\ref{mmol}), and the halo mass threshold under the environmental effects of LWB and SM (densely dashed, Eq.~\ref{mmol_ext})}
\label{fig:mth}       
\end{figure}

The results for $f_{\rm PBH}=10^{-4}$ (solid), $10^{-3}$ (dashed), 0.01 (dashed-dotted), 0.1 (dotted), $10^{-5}$ (long-dashed), and $10^{-6}$ (dot-dash-dotted) are shown in Fig.~\ref{fig:mth}, compared with the standard $\Lambda$CDM case (thick solid). 
As expected, $M_{\rm mol}$ increases with $f_{\rm PBH}$, by up to a factor of $\sim 3-5$ at $z\sim 20-40$ for $f_{\rm PBH}=0.1$, compared with the $\Lambda$CDM case (thick solid). The effect of PBHs is stronger at higher $z$ and the mass threshold converges to the $\Lambda$CDM value at low $z$. These results are generally consistent (within a factor of 2) with the cosmological simulations in \cite{Liu:2022okz}. Interestingly, we find that even for the extreme model with $f_{\rm PBH}=0.1$, the PBH-enhanced halo mass threshold without environmental effects remains below that with environmental effects of LWB and SM (Eq.~\ref{mmol_ext}). This indicates that the heating effect of PBHs on $M_{\rm mol}$ is likely unimportant in most regions of the universe as it is overwhelmed by the effects of LWB and SM. In the above analysis, we have ignored PDM substructures around PBHs which can reduce the central concentration of PDM and delay cloud collapse (see Sec.~\ref{sec:halo}), leading to higher $M_{\rm mol}$. This (dynamical) effect is only important for $f_{\rm PBH}\gtrsim 0.01$ according to \cite{Liu:2022okz}, and remains minor compared with the environmental effects of LWB and SM (Eq.~\ref{mmol_ext}). 
Here, we have also ignored the local LW radiation from accreting PBHs, which can potentially reduce $x_{\rm H_{2}}$ and increase $M_{\rm mol}$. This effect is sensitive to the spectra of accreting PBHs and gas densities around PBHs and may even be stronger than the PBH heating and environmental effects (see Sec.~\ref{sec:dcbh}). {Besides, the X-rays from accreting PBHs also regulate the thermochemistry of primordial gas via ionization and heating, which can either facilitate or delay Pop~III star formation (e.g., \cite{Hummel:2014sva,Ricotti:2016oyp,Park:2021sbs}).} Future studies should employ (radiative) hydrodynamic simulations to thoroughly assess the impact of local {LW and X-ray} feedback from PBHs on first star formation.

Finally, we combine the halo mass function $dn_{\rm h}/dM_{\rm h}$ and mass threshold $M_{\rm mol}$ of Pop~III star formation (as functions of $f_{\rm PBH}$ and $z$) to estimate the cosmic mass density of Pop~III stars ever formed. For simplicity, we assume that gas is converted into stars at a constant star formation efficiency $\epsilon_{\star,\rm III}$ in Pop~III star-forming halos with masses $M_{\rm h}\in [M_{\rm mol},10^{8}\ \rm M_{\odot}]$, independent of PBH parameters (which is justified in Sec.~\ref{sec:halo}), so the cosmic (co-moving) mass density of Pop~III stars (ever formed) can be written as
\begin{align}
    \rho_{\star,\rm III}=\epsilon_{\star,\rm III}\frac{\Omega_{b}}{\Omega_{m}}\int_{M_{\rm mol}}^{10^{8}\ \rm M_{\odot}}\frac{dn_{\rm h}}{dM_{\rm h}}M_{\rm h}dM_{\rm h}\ .\label{smd}
\end{align}
Since we are concerned with the difference made by PBHs, we focus on the ratio of $\rho_{\star,\rm III}$ between the PBH and standard $\Lambda$CDM models, i.e., $\rho_{\star,\rm III,PBH}/\rho_{\star,\rm III,CDM}$. 

\begin{figure}
\includegraphics[width=0.5\columnwidth]{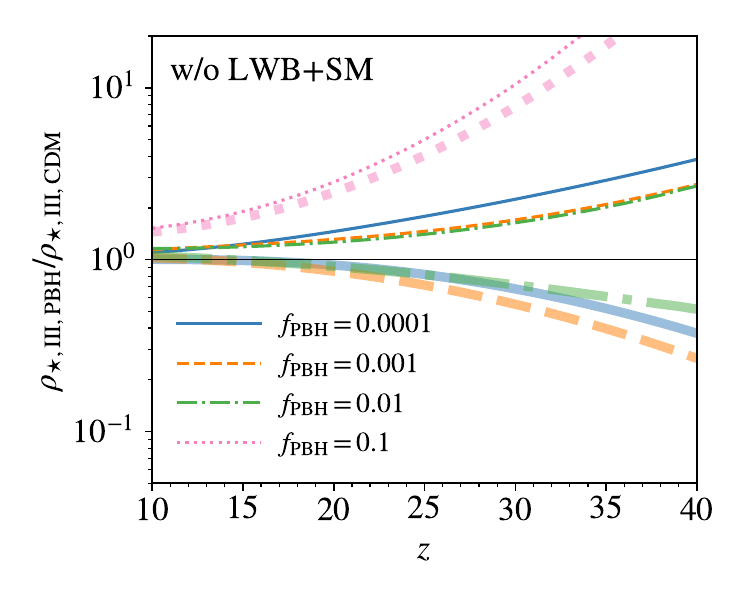} \includegraphics[width=0.5\columnwidth]{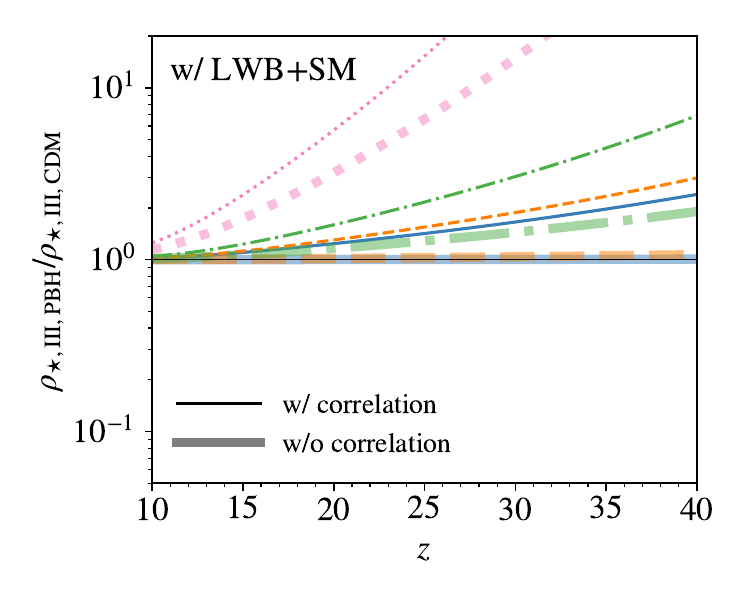}
\caption{Ratios of the mass densities of Pop~III stars ever formed in halos of $M_{\rm h}\in [M_{\rm mol},10^{8}\ \rm M_{\odot}]$ between the PBH and standard $\Lambda$CDM cases, for the PBH models with $m_{\rm PBH}=33\ \rm M_{\odot}$, $f_{\rm PBH}=10^{-4}$ (solid), $10^{-3}$ (dashed), 0.01 (dashed-dotted) and 0.1 (dotted), where the \textit{left panel} shows the results for the halo mass threshold $M_{\rm mol}$ without the environmental effects (see Fig.~\ref{fig:mth}), while the \textit{right panel} shows the results with the typical environmental effects of LWB and SM (Eq.~\ref{mmol_ext}), and the results with and without the heuristic correlation term (Eq.~\ref{pk_corr}) in the calculation of the halo mass function are shown for each PBH model with the thin and thick curves, respectively} 
\label{fig:fcol}       
\end{figure}

Fig.~\ref{fig:fcol} displays the results for $f_{\rm PBH}=10^{-4}$ (solid), $10^{-3}$ (dashed), 0.01 (dashed-dotted), and 0.1 (dotted), where we consider 2 models of $dn_{\rm h}/dM_{\rm h}$ with and without the correlation term in the linear power spectrum (Eq.~\ref{pk_corr}), denoted by the thin and thick curves, respectively. On the left panel, we show the results for the idealized case of $M_{\rm mol}$ without environmental effects (i.e., considering only the PBH heating effect on $M_{\rm mol}$). It turns out that for $f_{\rm PBH}\sim 10^{-4}-0.01$, $\rho_{\star,\rm III}$ is reduced by PBHs for the conservative estimate of $dn_{\rm h}/dM_{\rm h}$ without the correlation term in the linear power spectrum (Eq.~\ref{pk_corr}) by up to a factor $\sim 2$ at $z\lesssim 30$ (when most Pop~III stars form). On the contrary, given the optimistic estimate of $dn_{\rm h}/dM_{\rm h}$ with the correlation term, $\rho_{\star,\rm III}$ is increased by up to a factor $\sim 2$. In both cases, the effects of PBHs are stronger at higher $z$ and show weak, non-monotonic dependence on $f_{\rm PBH}$. On the right panel, we show the results for the mass threshold $M_{\rm mol}$ under the influence of typical environmental effects of LWB and SM which overwhelm the heating effect of PBHs. Here the effects of PBH are negligible for $f_{\rm PBH}\lesssim 10^{-3}$ without the correlation term, and $\rho_{\star,\rm III}$ increases monotonically with $f_{\rm PBH}$, reaching up to $\sim 3$ times the $\Lambda$CDM value at $z\lesssim 30$ for $f_{\rm PBH}\sim 10^{-4}-0.01$ when the optimistic estimate of $dn_{\rm h}/dM_{\rm h}$ is adopted. For all the 4 combinations of $M_{\rm mol}$ and $dn_{\rm h}/dM_{\rm h}$ models, $\rho_{\star,\rm III}$ is significantly enhanced by more than one order of magnitude at $z\gtrsim 30$ in the extreme case of $f_{\rm PBH}=0.1$. On the other hand, for $f_{\rm PBH}\sim 10^{-4}-0.01$ allowed by existing constraints, the variation of $\rho_{\star,\rm III}$ caused by PBHs is within a factor of 3, much smaller than the large ($\sim 1.5$~dex) uncertainties in current theoretical predictions on the cosmic Pop~III star formation history in the standard $\Lambda$CDM cosmology (see, e.g., Fig.~5 in \cite{Liu:2020krj}). Therefore, we conclude that the global impact of stellar-mass PBHs allowed by existing observational constraints on Pop~III star formation is minor.


\subsection{Internal structure of Pop~III star-forming halos and clouds}\label{sec:halo}
Although it is shown by the semi-analytical analysis above that the cosmic mass density of Pop~III stars is not changed greatly by stellar-mass PBHs with $f_{\rm PBH}\lesssim 0.01$, such PBHs may still have interesting local effects on the internal structure of star-forming halos and clouds. 
In this subsection, we use the results from the high-resolution cosmological zoom-in simulations in \cite{Liu:2022okz} as examples to demonstrate the local effects of PBHs at the halo and cloud scales. In particular, we consider the simulations for $f_{\rm PBH}=0$ (CDM), 0.001, 0.01, and 0.1 (with $m_{\rm PBH}=33\ \rm M_\odot$) in their Case~A zoom-in region with a mass resolution for PDM (gas) of $\sim 2\ (0.4)\ \rm M_\odot$. The runs start at $a_{\rm ini}=1/301$ from cosmological initial conditions generated by {a method similar to that}\footnote{Since the gravity of PBHs outside the zoom-in region is not considered, the enhancement of PBH perturbations (at large scales) by PBH clustering and mode correlation can be overestimated if all PBHs in the zoom-in region are used to evaluate the acceleration field (Eq.~\ref{acceleration}). Therefore, \cite{Liu:2022okz} restrict PBH perturbations to small scales by using at most the 64 nearest PBH particles within $2d_{\rm PBH}$ around the PDM/gas particle when evaluating Eq.~\ref{acceleration}, where $d_{\rm PBH}$ is the average separation between PBHs. Besides, $D_{\rm PBH}(a_{\rm ini})=T_{\rm iso}(a_{\rm ini})-1=D(a_{\rm ini})-1$ is used in \cite{Liu:2022okz}, without the correction motivated by the `seed' effect (see Sec.~\ref{sec:ic}), so the PBH perturbations in their simulations may be overestimated. Nevertheless, it is verified in \cite{Liu:2022okz} (see their Appendix A) that the conclusions made in this section on the internal structure are still valid despite such complications.} described in Sec.~\ref{sec:ic} with the resolution-based truncation (Eq.~\ref{trunc_grid}), suitable for the `Poisson' effect that is relevant for Pop~III star formation. The initial adiabatic density perturbations in the initial conditions are enhanced by adopting $\sigma_{8}=2$ rather than the cosmic mean $\sigma_{8}=0.8159$ \cite{Planck:2015fie} to mimic the typical overdense regions where the first stars form at $z\sim 30$. These simulations are also equipped with a non-equilibrium primordial chemistry and cooling network and sub-grid models for BH accretion, dynamical friction and feedback (which is consistent with the model in Sec.~\ref{sec:accretion}) to self-consistently capture the gravitational and heating effects of PBHs on the nonlinear dynamics of halo assembly and cloud collapse as well as the thermal and chemical evolution of primordial gas. The reader is referred to Sec.~2 of \cite{Liu:2022okz} for details on the numerical methods and setup of the simulations\footnote{In fact, when environmental effects are ignored, first star formation is delayed for $f_{\rm PBH}\gtrsim 0.01$ in the zoom-in simulations from \cite{Liu:2022okz}, which seems in conflict with the prediction by Eq.~\ref{smd} using the optimistic estimate of the halo mass function from the PS formalism with the correction term (Eq.~\ref{pk_corr}). On the other hand, star formation is accelerated in the zoom-in simulations for $f_{\rm PBH}\lesssim 0.01$, inconsistent with the semi-analytical prediction without the correction term. It is difficult to interpret these discrepancies as the trends obtained from zoom-in simulations are sensitive to the specific halo assembly histories in the small, biased zoom-in regions (i.e., cosmic variance) and cannot be naively generalized. Cosmological simulations in larger volumes are required to validate/calibrate the PS formalism (see \cite{Zhang:2024ytf}).}. 

\begin{figure}
\includegraphics[width=1\columnwidth]{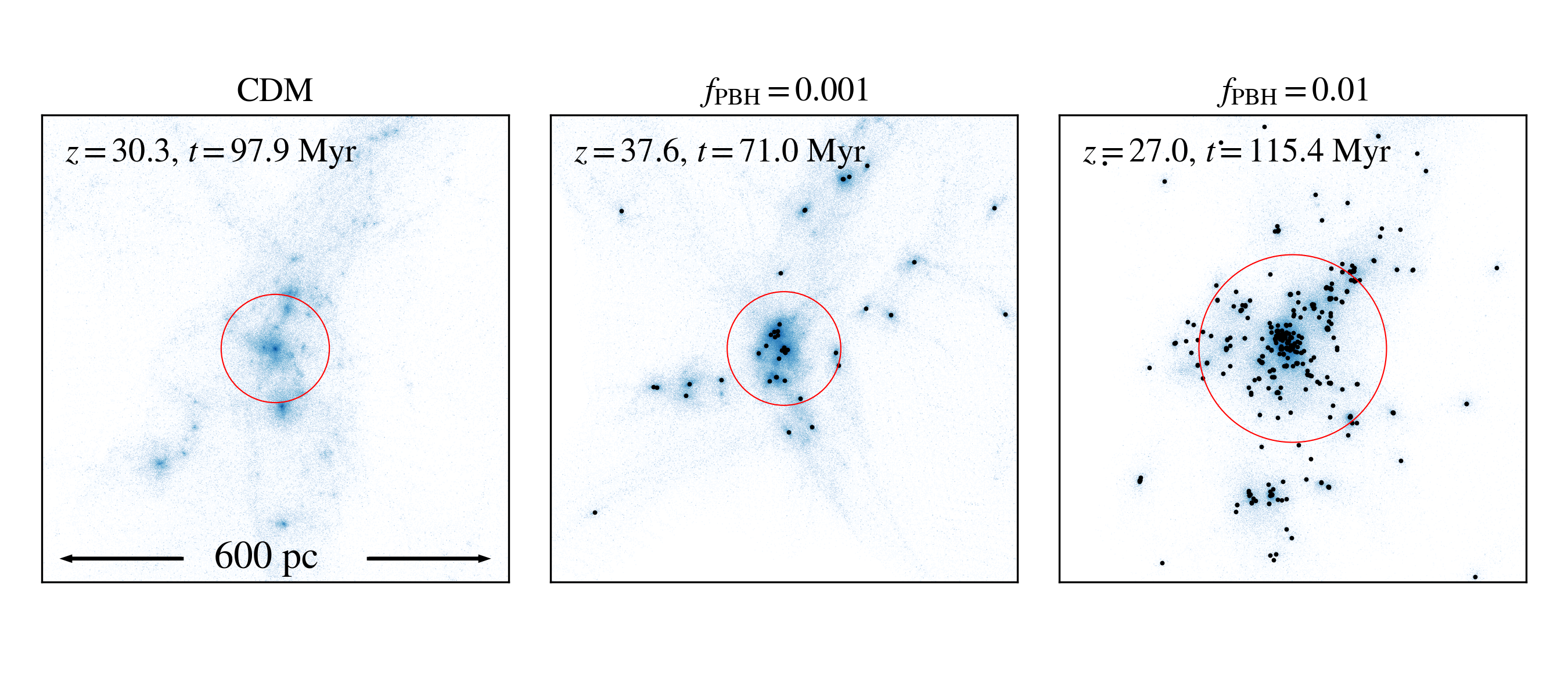}
\caption{Projected distributions of PDM and PBHs at the onset of star formation in a data slice with a (physical) extent of $600\ \rm pc$ and a thickness of $300\ \rm pc$, for the reference $\Lambda$CDM simulation (\textit{left}) and PBH simulations with $m_{\rm PBH}=33\ \rm M_{\odot}$, $f_{\rm PBH}=0.001$ (\textit{middle}) and $0.01$ (\textit{right}), in the Case A zoom-in region from \cite{Liu:2022okz}, where PBHs are plotted with black dots, and the circles indicate the halo virial radii}
\label{fig:mass_dis}       
\end{figure}

We focus on the snapshot when the maximum hydrogen number density reaches $10^{5}\ \rm cm^{-3}$, which marks the onset of Pop~III star formation. At this moment, a dense ($n_{\rm H}\gtrsim 10^{4}\ \rm cm^{-3}$) cold ($T\lesssim 10^{3}\ \rm K$) gas clump of $\sim 10^{3}\ \rm M_{\odot}$ has formed by run-away collapse with efficient molecular cooling in the central parsec of the most massive halo in each simulation, which is the typical environment for Pop~III star formation. Such a dense gas clump is expected to further collapse and form Pop~III stars in a very short (free-fall) timescale of $t_{\rm ff}\sim 0.5$~Myr. The simulations stop at this moment because higher resolution is required to follow the subsequent collapse. The fact that such typical Pop~III star-forming gas clumps occur in all simulations with different PBH parameters implies that the standard picture of Pop~III star formation at sub-parsec scales is not changed by the stellar-mass PBHs considered here. Nevertheless, PBHs can have noticeable effects in the outer regions of the halo via their heating effect and the PDM substructures around them, as shown in Fig.~\ref{fig:mass_dis}. 

\begin{figure}
\includegraphics[width=1\columnwidth]{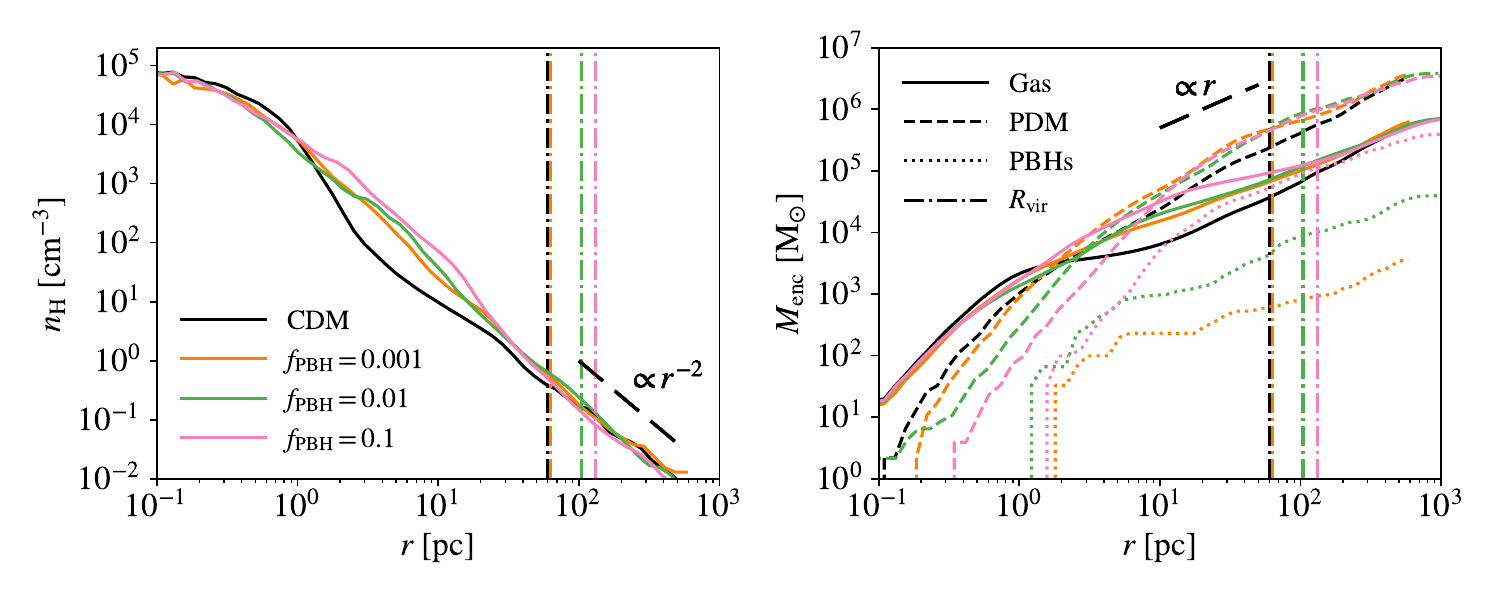}
\caption{Internal structure of the star-forming halo at the onset of star formation in terms of the hydrogen number density profile (\textit{left}) and enclosed mass profiles (\textit{right}) for gas (solid), PDM (dashed) and PBHs (dotted), for the reference $\Lambda$CDM (CDM) simulation (black) and PBH simulations with $m_{\rm PBH}=33\ \rm M_{\odot}$, $f_{\rm PBH}=0.001$ (orange), $0.01$ (green), and 0.1 (pink), in the Case A zoom-in region from \cite{Liu:2022okz}, where the halo virial radii are shown with the vertical dashed-dotted lines}
\label{fig:denspro}       
\end{figure}

To better appreciate this point, we plot the density profiles of the halos hosting the Pop~III star-forming clumps in Fig.~\ref{fig:denspro} for $f_{\rm PBH}=0$ (CDM, black), $0.001$ (orange), $0.01$ (green), and 0.1 (pink). Here the halo/cloud center is defined as the gas density peak. The left panel shows the hydrogen number density profiles $n_{\rm H}(r)$, and the right panel shows the enclosed mass profiles $M_{\rm enc}(r)$ of all components: gas (solid), PDM (dashed), and PBHs (dotted). We find that indeed the gas density profiles are almost identical in the central region ($\lesssim 1$~pc) occupied by the star-forming gas clump for all simulations. As $f_{\rm PBH}$ increases, the gas distribution in the outer part of the halo ($1\ \mathrm{pc}\lesssim r\lesssim R_{\rm vir}$) becomes more clumpy, and {the density is higher for $r\sim 1-20\ \rm pc$, showing that gas is pilled up in this region}. This is likely caused by the PBH heating and PDM substructures around PBHs that can slow down cloud collapse. Nevertheless, as $r\rightarrow R_{\rm vir}$, the gas density profile converges to the power-law form, $n_{\rm H}(r)\propto r^{-2}$, of an isothermal distribution, which is a typical feature of Pop~III star-forming halos (see, e.g., \cite{Hirano:2015wxa}). The density profile of PDM also becomes shallower with higher $f_{\rm PBH}$, especially in the central region ($r\lesssim 10\ \rm pc$) where the PDM density decreases with $f_{\rm PBH}$. This indicates that it is more difficult for PDM to concentrate at the center when there are more PBHs, consistent with the results in \cite{Inman:2019wvr}. The reason is that substructures of PDM around PBHs are more tightly bound and therefore more difficult to destroy/strip during virialization compared with their BH-less counterparts in the $\Lambda$CDM case (see Fig.~\ref{fig:mass_dis}). For PBHs, we have $M_{\rm enc}(r)\propto r$ at $r\gtrsim 10$~pc, corresponding to an isothermal distribution independent of $f_{\rm PBH}$, while the density profile is shallower in the central region with increasing $f_{\rm PBH}$. In fact, only up to a few PBHs can reach $r\lesssim 2\ \rm pc$ regardless of the overall PBH abundance, and there are no PBHs at $r<1$~pc in all the 19 simulations run by \cite{Liu:2022okz} that include PBH heating. This interesting outcome can be interpreted with the survivor bias for gas condensation. Since we define the halo center as the gas density peak, PBHs cannot be too close to the center, because otherwise, gas will not be able to cool and condense due to the strong heating of nearby PBHs. 

\begin{figure}
\includegraphics[width=1\columnwidth]{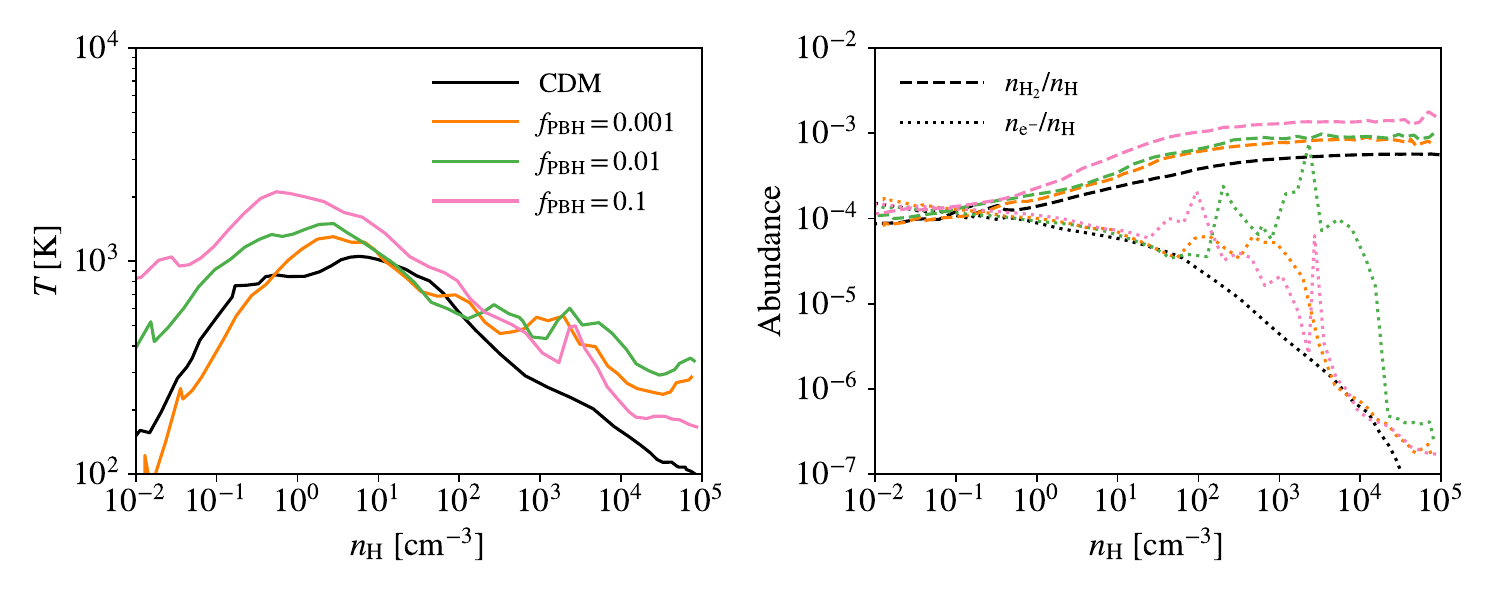}
\caption{Thermal and chemical properties of the star-forming halo at the onset of star formation in terms of the temperature-density diagram (\textit{left}) and abundances of $\rm H_{2}$ (dashed) and $\rm e^{-}$ (dotted) as functions of density (\textit{right}), for the reference $\Lambda$CDM (CDM) simulation (black) and PBH simulations with $m_{\rm PBH}=33\ \rm M_{\odot}$, $f_{\rm PBH}=0.001$ (orange), $0.01$ (green), and 0.1 (pink), in the Case A zoom-in region from \cite{Liu:2022okz}}
\label{fig:cpro}       
\end{figure}

Next, we evaluate the impact of PBHs on the thermal and chemical properties of gas in Fig.~\ref{fig:cpro}, where we show the temperature-density behavior (left) and abundances of $\rm H_{2}$ (dashed) and free electrons $\rm e^{-}$ (dotted) as functions of density (right) for the same $\Lambda$CDM and PBH models considered in Fig.~\ref{fig:denspro}. 
{The gas temperature generally increases with $f_{\rm PBH}$ for $f_{\rm PBH}\gtrsim 0.001$ at halo outskirts with $n_{\rm H}\lesssim 1\ \rm cm^{-3}$ where cooling is relatively inefficient. The reason is that the halo mass at the onset of star formation increases with $f_{\rm PBH}$ due to the heating of PBHs, which results in stronger compression heating by halo assembly with a higher virial temperature. }
The $\rm H_{2}$ abundance also increases with $f_{\rm PBH}$ by a factor almost independent of density at $n_{\rm H}\gtrsim 10\ \rm cm^{-3}$, as the heating and ionization from accreting PBHs facilitate $\rm H_{2}$ formation. This trend can be described by a simple power-law fit in Eq.~\ref{xh2}. Beyond these clear trends with $f_{\rm PBH}$, we find large temporal variations of temperature and electron abundance as functions of density in the central region with $n_{\rm H}\gtrsim 10^{2}\ \rm cm^{-3}$ (i.e., $r\lesssim 10\ \rm pc$, see the left panel of Fig.~\ref{fig:denspro}), although they are always above the standard $\Lambda$CDM evolution track. 
This shows that gas properties are sensitive to the detailed stochastic gaseous environments around a small number of PBHs in the central region. 

Finally, let us consider the fate of the dense clumps formed at the end of the simulations under the influence of PBHs. These clumps are expected to collapse and form Pop~III stars within a timescale of $t_{\rm ff}\sim 0.5$~Myr, while nearby PBHs will sink to the clump center over the dynamical friction timescale $t_{\rm DF}$.
If a PBH can reach the densest central region ($r\lesssim 10^{-3}$~pc) with $n_{\rm H}\sim 10^{9}\ \rm cm^{-3}$ before completion of star formation (and evaporation of the cloud by stellar feedback), it can undergo super-Eddington accretion to grow significantly (i.e., by a factor of $\sim 10$ within $\sim 0.01$~Myr) and even suppress star formation by {evaporating} the cloud with accretion feedback (see Sec. 4.3 in \cite{Liu:2022okz}). 
We can compare the dynamical friction timescale $t_{\rm DF}$ of PBHs with the cloud collapse timescale $t_{\rm ff}$ to discern whether PBHs can reach the center to significantly affect the star formation process. For a BH of a mass $m_{\rm BH}$ with an initial apocentric distance $r$ and a velocity $v_{\rm BH}$ with respect to the center, the dynamical friction timescale can be estimated by \cite{binney2011galactic}
\begin{align}
    \frac{t_{\rm DF}}{\mathrm{Myr}}\simeq \frac{340}{\ln\Lambda}\left(\frac{r}{\rm3\ pc}\right)^{2}\left(\frac{v_{\rm BH}}{\rm 10\ km\ s^{-1}}\right)\left(\frac{m_{\rm BH}}{100\ \rm M_{\odot}}\right)^{-1}\ , \label{tdf}
\end{align}
where $\ln\Lambda\sim 10$ is the Coulomb logarithm. In our case of $m_{\rm BH}= 33\ \rm M_{\odot}$, the simulations predict that $v_{\rm BH}\sim 5-10\ \rm km\ s^{-1}$ and $r\gtrsim 1\ \rm pc$ (due to the survivor bias of gas condensation), so we have $t_{\rm DF}\gtrsim 6\ {\rm Myr}\gg t_{\rm SF}$. This indicates that stellar-mass PBHs are unlikely to sink into Pop~III star-forming disks and, therefore, cannot { interact with Pop~III protostars directly}. 
Nevertheless, $t_{\rm DF}$ is comparable to the typical lifetimes of Pop~III stars and relaxation timescales of Pop~III star clusters, so PBHs may still affect Pop~III stellar evolution and dynamics of Pop~III star clusters, which can be an interesting topic for future research. 

\subsection{Impact of PBHs on the formation of massive BH seeds}\label{sec:dcbh}

We have shown that the impact of stellar-mass PBHs (allowed by existing observational constraints) on Pop~III star formation in molecular-cooling minihalos ($M_{\rm h}\sim 10^{5}-10^{8}\ \rm M_{\odot}$) is likely minor both globally and locally. However, the effects of PBHs may still be important in larger, atomic-cooling halos ($M_{\rm h}\gtrsim 10^{8}\ \rm M_\odot$). In particular, if PBH accretion produces strong LW radiation that can efficiently dissociate $\rm H_{2}$ in a metal-pool halo, molecular cooling and fragmentation will be suppressed, such that gas can collapse directly via atomic cooling to form a massive ($\sim 10^{4}-10^{6}\ \rm M_{\odot}$) BH \cite{Bromm:2003}. This direct-collapse BH (DCBH) scenario (reviewed by, e.g., \cite{Latif:2018fzc,Haemmerle:2020iqg}) is an important channel of forming the seeds of the supermassive BHs observed in high-$z$ quasars, but it is only expected to occur in very rare sites with strong external LW radiation fields or high inflow rates in the standard $\Lambda$CDM cosmology \cite{Visbal:2014yga,Wise2019,Regan:2019vdf}. 
To explore whether DCBH formation can be enhanced by PBHs, we calculate the LW intensity produced by PBHs at the halo center, $J_{21,\rm PBH}(r=0)$. Here we treat the LW radiation from PBHs as an `external' background by ignoring self-shielding, in order to compare our results with the typical critical intensity of DCBH formation $J_{21,\rm crit}= 1000$, defined for an external radiation field with a temperature $T_{\rm rad}\gtrsim 2.5\times 10^{4}\ \rm K$ \cite{Sugimura:2014sqa}. We assume that both gas and PBHs follow isothermal distributions ($M_{\rm enc}\propto r$), based on simulation results \cite{Wise2019,Safarzadeh:2020vbv,Liu:2022okz}. 
The PBH distribution is further truncated at $r=1$~pc considering the survivor bias of gas condensation. We use Eq.~\ref{bondi} to calculate the accretion rate of each PBH given the characteristic velocity $\tilde{v}\sim \sqrt{GM_{\rm h}/R_{\rm vir}}$, and derive the specific luminosity in the LW band from the accretion rate using the spectrum model in \cite{Takhistov:2021aqx}. We integrate over the volume defined by the radius range $r\in [1\ {\rm pc},R_{\rm vir}]$ to sum up the contributions from individual PBHs. 

\begin{figure}
\sidecaption
\includegraphics[width=0.7\columnwidth]{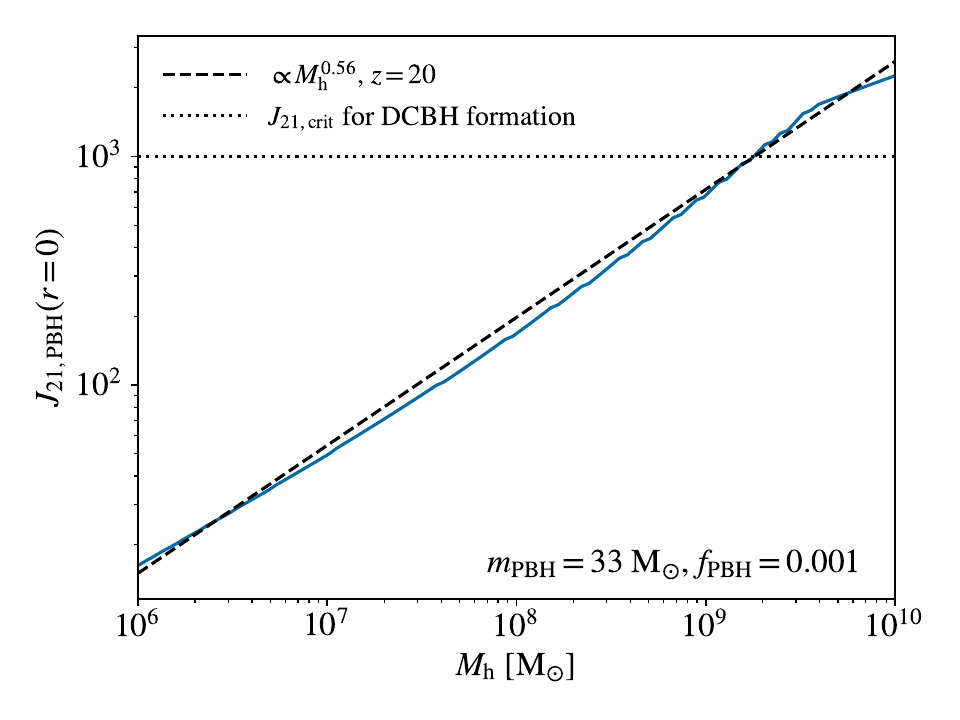}
\caption{Intensity of the LW radiation at the halo center from accreting PBHs with $m_{\rm PBH}=33\ \rm M_{\odot}$ and $f_{\rm PBH}=0.001$ as a function of halo mass for $z=20${, assuming isothermal distributions of gas and PBHs}, ignoring self-shielding (i.e., treating the LW radiation as an `external' background), where the dashed line shows the power-law fit (Eq.~\ref{j21_pbh}) that captures the results of detailed calculations (solid) within a factor of 2 uncertainty, for $m_{\rm PBH}\sim 10-100\ \rm M_{\odot}$ and $M_{\rm h}\sim 10^{6}-10^{10}\ \rm M_{\odot}$ at $z\sim 10-30$, and the dotted horizontal line denotes the critical intensity, $J_{21,\rm crit}\sim 1000$, for DCBH formation \cite{Sugimura:2014sqa}}
\label{fig:j21}       
\end{figure}

The resulting $J_{21,\rm PBH}(r=0)$ as a function of halo mass at $z=20$ for $f_{\rm PBH}=0.001$ is shown in Fig.~\ref{fig:j21} as an example. In fact, we find that our results can be described by a simple power-law fit, within a factor 2 uncertainty, for $m_{\rm PBH}\sim 10-100\ \rm M_{\odot}$ and $M_{\rm h}\sim 10^{6}-10^{10}\ \rm M_{\odot}$ at $z\sim 10-30$:
\begin{align}
    J_{21,\rm PBH}(r=0)\sim 5\times 10^{4}f_{\rm PBH}(M_{\rm h}/10^{7}{\ \rm M_{\odot}})^{0.56}(m_{\rm PBH}/{33\ \rm M_{\odot}})^{0.9}\ .\label{j21_pbh}
\end{align}
This implies that halos with $M_{\rm h}\gtrsim 3.3\times 10^{7}{\ \rm M_{\odot}}\left( f_{\rm PBH}/0.01\right)^{-1.8}(m_{\rm PBH}/{33\ \rm M_{\odot}})^{-1.6}$ will meet the criterion $J_{21}\gtrsim 10^{3}$ for DCBH formation \cite{Sugimura:2014sqa}. Therefore, for $f_{\rm PBH}\sim 0.001-0.01$ still (marginally) allowed by observational constraints, stellar-mass PBHs can potentially trigger the formation of massive BH seeds via direct collapse in typical atomic-cooling halos ($M_{\rm h}\sim 10^{8}-10^{10}\ \rm M_\odot$). Radiative hydrodynamic simulations with detailed modeling of the radiation fields from accreting PBHs are required to verify this possibility.

\section{Formation of massive galaxies seeded by SMPBHs}
\label{sec:smpbh}
JWST has recently discovered a population of unusually massive (stellar masses $\gtrsim 10^{10}\ \rm M_{\odot}$) galaxy candidates formed at $z\gtrsim 8$ \cite{Carnall2023,Labbe2023,Glazebrook:2023vkx,Carniani2024,deGraaff2024,Lopez-Corredoira:2024pgl,UrbanoStawinski2024,Nanayakkara2024} and a surprising ubiquity of (overmassive) SMBHs in galaxies at $z\gtrsim 5$ \cite{Goulding:2023gqa,Harikane2023,Greene:2024phl,Maiolino:2023bpi,Matthee:2023utn,Natarajan:2023rxq,Akins2024,Durodola:2024bom}, which is seemingly in tension with typical theoretical predictions on high-$z$ galaxy and SMBH formation in the standard $\Lambda$CDM cosmology \cite{Boylan-Kolchin:2022kae,Menci:2022wia,Chen:2023ugq}. 
A promising explanation is SMPBHs (or other primordial massive objects like cosmic string loops \cite{Jiao:2023wcn}) 
as a component of DM, which can serve as seeds of both SMBHs and massive galaxies \cite{Liu:2022bvr,Su:2023jno,Colazo:2024jmz}. Such enhanced early galaxy formation by SMPBHs may also be able to explain the excess of galaxy-galaxy strong lensing probability and star formation observed in (proto-)clusters of galaxies \cite{Remus:2022hxb,Meneghetti:2023fug}, and mergers of SMPBHs can also contribute to the SGWB recently discovered by pulsar timing arrays \cite{Gouttenoire:2023nzr} (see Part IV of the book). In this section, we use the extraordinarily massive old galaxy candidate ZF-UDS-7329 detected by JWST (with a spectroscopic redshift $z=3.21$) \cite{Glazebrook:2023vkx} as an example to demonstrate how the SMPBH parameters required to explain certain observations can be identified with a back-of-the-envelope calculation focusing on the `seed' effect {(see also \cite{deGraaff2024,UrbanoStawinski2024,Nanayakkara2024} for JWST observations of similar massive old galaxy candidates unexpected in the standard $\Lambda$CDM cosmology). 

It is found by \cite{Glazebrook:2023vkx} that the observed spectrum of ZF-UDS-7329 can be explained by an old stellar population formed at $z\gtrsim 11$ with a total stellar mass of $M_{\star}\sim 2.5\times 10^{11}\ \rm M_\odot$. The co-moving number density of galaxies like ZF-UDS-7329 is inferred as $n_{\rm g}\sim 6\times 10^{-7}\ \rm Mpc^{-3}$ from the survey volume. In the standard $\Lambda$CDM cosmology, such massive galaxies (and the corresponding massive halos) are not expected to occur with the inferred abundance until $z\sim 6$. Now, we assume that this galaxy resides in a halo seeded by an SMPBH with an initial mass of $m_{\rm PBH}$, such that the halo mass follows $M_{\rm h}\sim M_{\rm B}(m_{\rm PBH},z)=m_{\rm PBH}/[(1+z)a_{\rm eq}]$ in the idealized case of isolated growth. To reproduce the observed stellar mass and abundance of ZF-UDS-7329, we simply require $\bar{n}_{\rm PBH}>n_{\rm g}$ and {$M_{\star}=\epsilon (\Omega_{b}/\Omega_{m})M_{\rm B}$} at $z=11$, given the star formation efficiency $\epsilon<1$ and the average number density of SMPBHs $\bar{n}_{\rm PBH}=f_{\rm PBH}[3H_{0}^{2}(\Omega_{m}-\Omega_{b})]/(8\pi Gm_{\rm PBH})$. Considering the typical (extreme) limit $\epsilon< 0.1$ (1) for such massive galaxies, we have {$m_{\rm PBH}\gtrsim 5.6\times 10^{10\ (9)}\ \rm M_\odot$ and $f_{\rm PBH}\gtrsim 10^{-6\ (7)}$}. We also require $f_{\rm PBH}<(1+z)a_{\rm eq}\sim 0.0035$, which is a necessary (but not sufficient) condition for isolated growth of PBH seeded halos, and $m_{\rm PBH}\lesssim 10^{11}\ \rm M_\odot$ considering the non-detection of BHs above $10^{11}\ \rm M_{\odot}$. We summarize these requirements and the resulting region in the $m_{\rm PBH}$-$f_{\rm PBH}$ parameter space in Fig.~\ref{fig:seed}.

\begin{figure}
\sidecaption
\includegraphics[width=0.7\columnwidth]{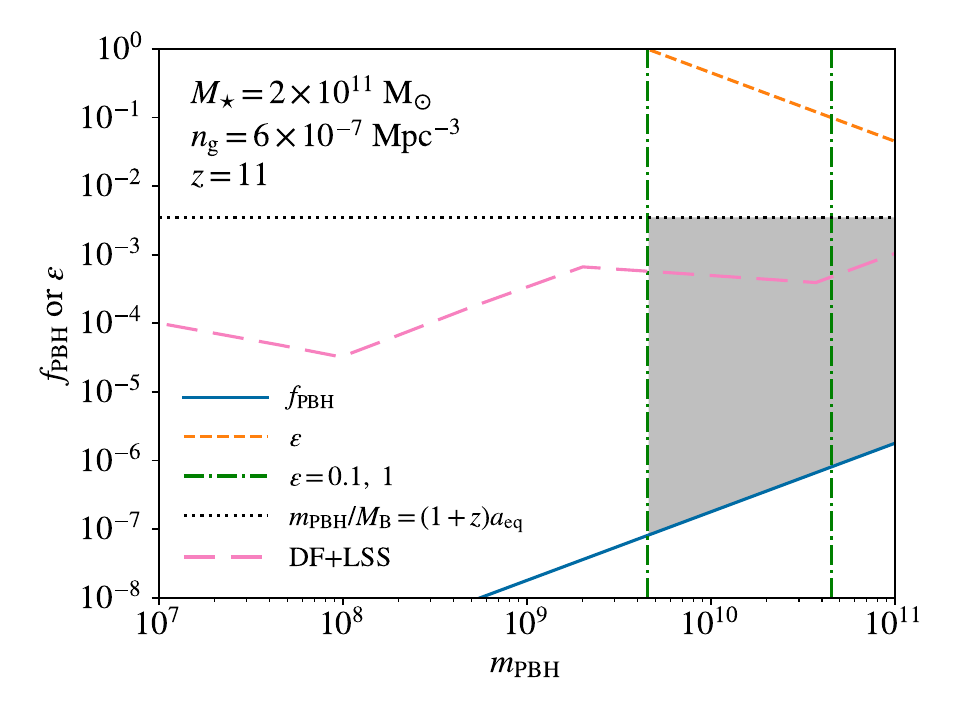}
\caption{The PBH parameters (shaded region) required to explain the extraordinarily massive galaxy candidate ZF-UDS-7329 formed at $z=11$ \cite{Glazebrook:2023vkx} with the `seed' effect of SMPBHs, where the solid line shows the lower limit of $f_{\rm PBH}$, given by the galaxy abundance requirement $\bar{n}_{\rm PBH}>n_{\rm g}$, the dashed line shows the star formation efficiency $\epsilon$ required to explain the stellar mass with $M_{\star}=\epsilon M_{\rm B}$, the vertical dash-dotted lines label typical values of $\epsilon=0.1$ and 1, and the horizontal dotted lines denotes the necessary (but not sufficient) condition $f_{\rm PBH}<(1+z)a_{\rm eq}$ for isolated growth, and the long-dashed curve shows the combined constraint from dynamical friction (DF) \cite{Carr:1997cn} and large-scale structure (LSS) \cite{Carr:2018rid} compiled by \cite{Carr:2020gox}}
\label{fig:seed}       
\end{figure}

In fact, if PBHs form in the standard spherical collapse scenario, and primordial density fluctuations are Gaussian, such SMPBH models with $m_{\rm PBH}\sim 10^{10}\ \rm M_\odot$ and $f_{\rm PBH}\sim 10^{-7}-0.0035$ are ruled out by the observed limit of CMB $\mu$-distortion \cite{Carr:2020gox} (see also Part V of the book). 
However, the CMB $\mu$-distortion constraint can be weakened/lifted if primordial density fluctuations are highly non-Gaussian {\cite{Nakama:2017xvq,Hooper:2023nnl}} or PBHs form in non-standard scenarios such as inhomogeneous baryogenesis with the modified Affleck-Dine mechanism \cite{Kawasaki:2019iis,Kasai:2022vhq,Kasai:2024tgu}, {first-order phase transitions \cite{Jedamzik:1999am,Liu:2021svg,Davoudiasl:2021ijv}, and the collapse of supercritical vacuum bubbles nucleated during inflation \cite{Garriga:2015fdk,Huang2023jun}} (see Part II of the book). If we bypass the CMB $\mu$-distortion constraint with these `non-standard' scenarios, the strongest remaining constraints around this mass scale come from the infall of PBHs into the Galactic center by dynamical friction \cite{Carr:1997cn} and large-scale structure statistics \cite{Carr:2018rid}, which jointly require {$f_{\rm PBH}\lesssim 10^{-3}$} (see the long-dashed curve in Fig.~\ref{fig:seed})\footnote{The UV luminosity functions of galaxies observed by HST at $z\sim 8$ also place constraints on PBH parameters which, compared with the constraints from large-scale structure \cite{Carr:2018rid}, are slightly weaker in the conservative scenario of the `Poisson' effect modeled by the PS formalism with suppressed small-scale power at $k>k_{\star}$ from nonlinear dynamics around PBHs (see Fig.~1 in \cite{Gouttenoire:2023nzr}), and is thus not shown here. The constraints from HST observations can be stronger with more optimistic estimates of the small-scale power of PBH perturbations (see Fig.~3 in \cite{Gouttenoire:2023nzr}), but still unable to completely rule out SMPBHs as seeds of the observed massive galaxy candidates. }. In this case, a substantial region in the PBH parameter space ($m_{\rm PBH}\sim 10^{7}-10^{11}\ \rm M_\odot$, {$f_{\rm PBH}\sim 10^{-7}-10^{-3}$}) is still allowed to explain recent JWST observations \cite{Labbe2023,Glazebrook:2023vkx,Liu:2022bvr,Su:2023jno}. However, a key uncertainty in our idealized analysis is the star formation efficiency $\epsilon$ that captures the complex interplay between star formation and BH accretion feedback. It is possible that star formation will be significantly suppressed by the SMPBH, leading to extremely small $\epsilon$ that requires $m_{\rm PBH}\gtrsim 10^{11}\ \rm M_\odot$ in conflicts with observations. SMPBHs and the massive galaxies seeded by them may also violate other observational constraints from the early Universe, such as the 21 cm signal, ionization history, cosmic infrared and X-ray backgrounds \cite{Kashlinsky:2016}. Moreover, in the presence of PBH-seeded halos, first star formation can happen at very high redshifts ($z\gtrsim 100$) under peculiar conditions, leading to distinct stellar properties compared with the standard $\Lambda\rm CDM$ case \cite{Ito:2024mla}. Future studies with detailed modeling of the interactions between PBHs, baryons, and PDM (in cosmological simulations) are required to fully understand star/galaxy formation in PBH cosmologies and evaluate their viability against observations.

\section{Summary and Outlook}
\label{sec:dis}
A possible PBH contribution to the cosmic DM density adds qualitatively new aspects to the physics of cosmological structure formation. 
The first stars and galaxies are probing cosmic structure on the smallest scales, and provide therefore an ideal laboratory for the nature of DM, both the microphysics of PDM, reflecting extensions to the standard model of particle physics, as well as the mass spectrum and abundance of PBHs, similarly depending on exotic physical processes in the ultra-early Universe. 
As we have discussed in this chapter, the novel aspects from PBHs present challenges to existing analytical and simulation-based treatments of first star/galaxy formation. We have suggested idealized, or heuristic approaches that can get us started, but need to be further developed in future work. 
In particular, the following topics deserve further investigation:
\begin{itemize}
    \item {Halo mass functions in PBH cosmologies from robust analytical models and cosmological simulations that capture the interplay between the `Poisson' and `seed' effects of PBHs on structure formation}
    \item Effects of LW { and X-ray} feedback from PBHs on early star formation and DCBH formation triggered by PBHs
    \item Interplay between star formation, stellar feedback, BH accretion and feedback in structures induced by SMPBHs
\end{itemize}
The overall motivation for such dedicated effort is the importance of broadly exploring the behavior and viability of PBH dark matter, in terms of not violating existing and future empirical constraints.

Developing this astrophysical laboratory may indeed be the only way to conquer the dark matter frontier if the production or detection of dark matter particles continues to elude experimental physics. More specifically, for PBHs, the challenge is to disentangle their properties and impact on cosmic history from the suite of astrophysical black holes. Here, again, the very onset of star and galaxy formation promises to provide unique fingerprints for black holes of such vastly different origins. We thus face an exciting theoretical and computational challenge in elucidating PBH-$\Lambda$CDM cosmology, at the very moment when frontier telescopes, such as the JWST, Euclid, and soon the Roman Space Telescope, and other observational facilities are coming online.  



 \bibliographystyle{unsrt}
 \bibliography{authorsample.bib}

\end{document}